\documentclass[acmsmall, nonacm]{acmart}
\AtBeginDocument{%
  }

\setcopyright{acmlicensed}
\copyrightyear{2025}
\acmYear{2025}
\acmDOI{XXXXXXX.XXXXXXX}

\acmJournal{TOCHI}
\acmVolume{1}
\acmNumber{1}
\acmArticle{0}
\acmMonth{8}

\usepackage{booktabs}
\usepackage{multirow}
\usepackage{siunitx}

\begin{document}
\title{Audio Personas: Augmenting Social Perception via Body-Anchored Audio Cues}

\author{Yujie Tao}
\orcid{0000-0002-3492-6747}
\email{yjtao@stanford.edu}
\affiliation{%
  \institution{Stanford University}
  \city{Stanford}
  \state{California}
  \country{USA}
}

\author{Libby Ye}
\orcid{0009-0002-9385-5501}
\affiliation{%
 \institution{Stanford University}
 \city{Stanford}
 \state{California}
 \country{USA}}
 \email{libbyye@stanford.edu}

\author{Jeremy N. Bailenson}
\orcid{0000-0003-2813-3297}
\affiliation{%
  \institution{Stanford University}
  \city{Stanford}
  \state{California}
  \country{USA}}
  \email{bailenso@stanford.edu}

\author{Sean Follmer}
\orcid{0000-0001-5592-5949}
\affiliation{%
  \institution{Stanford University}
  \city{Stanford}
  \state{California}
  \country{USA}}
\email{sfollmer@stanford.edu}

\renewcommand{\shortauthors}{Tao et al.}

\begin{abstract}
  We introduce Audio Personas, enabling users to "decorate" themselves with body-anchored sounds in audio augmented reality. Like outfits, makeup, and fragrances, audio personas offer an alternative yet dynamic channel to augment face-to-face interactions. For instance, one can set their audio persona as rain sounds to reflect a bad mood, bee sounds to establish personal boundaries, or a playful "woosh" sound to mimic passing by someone like a breeze. To instantiate the concept, we implemented a headphone-based prototype with multi-user tracking and audio streaming. Our preregistered in-lab study with 64 participants showed that audio personas influenced how participants formed impressions. Individuals with positive audio personas were rated as more socially attractive, more likable, and less threatening than those with negative audio personas. Our study with audio designers revealed that audio personas were preferred in public and semi-public-private spaces for managing social impressions (e.g., personality) and signaling current states (e.g., emotions).\footnote{Manuscript accepted to ACM Transactions on Computer-Human Interaction (TOCHI). This is the authors' preprint version.}

\end{abstract}



\maketitle
\section{Introduction}
In digital environments, such as social media or virtual reality, users can transform how others perceive them by adjusting the characteristics of their digital avatars. With just a few clicks, one can modify attributes like hairstyle, facial features, and body type--or even embody completely different identities by switching between genders or adopting non-humanoid avatars~\cite{turkle2011life}. 

While people can easily modify their social images online, face-to-face interactions lack such flexibility. Physical adornments like makeup and clothing remain relatively static, limited in their ability to convey changing social messages. As such, there is a growing trend toward integrating digital elements, traditionally limited to virtual interactions, into real-world settings. One prominent example is the use of visual filters, popularized through social media, are now adopted in mobile augmented reality (mobile AR). By anchoring visual effects such as face masks or virtual jewelries to one's body, it alters how one is socially perceived by others~\cite{hakkila2017clothing, Roinesalo2016}.

\begin{figure}[hbt!]
  \includegraphics[width=\textwidth]{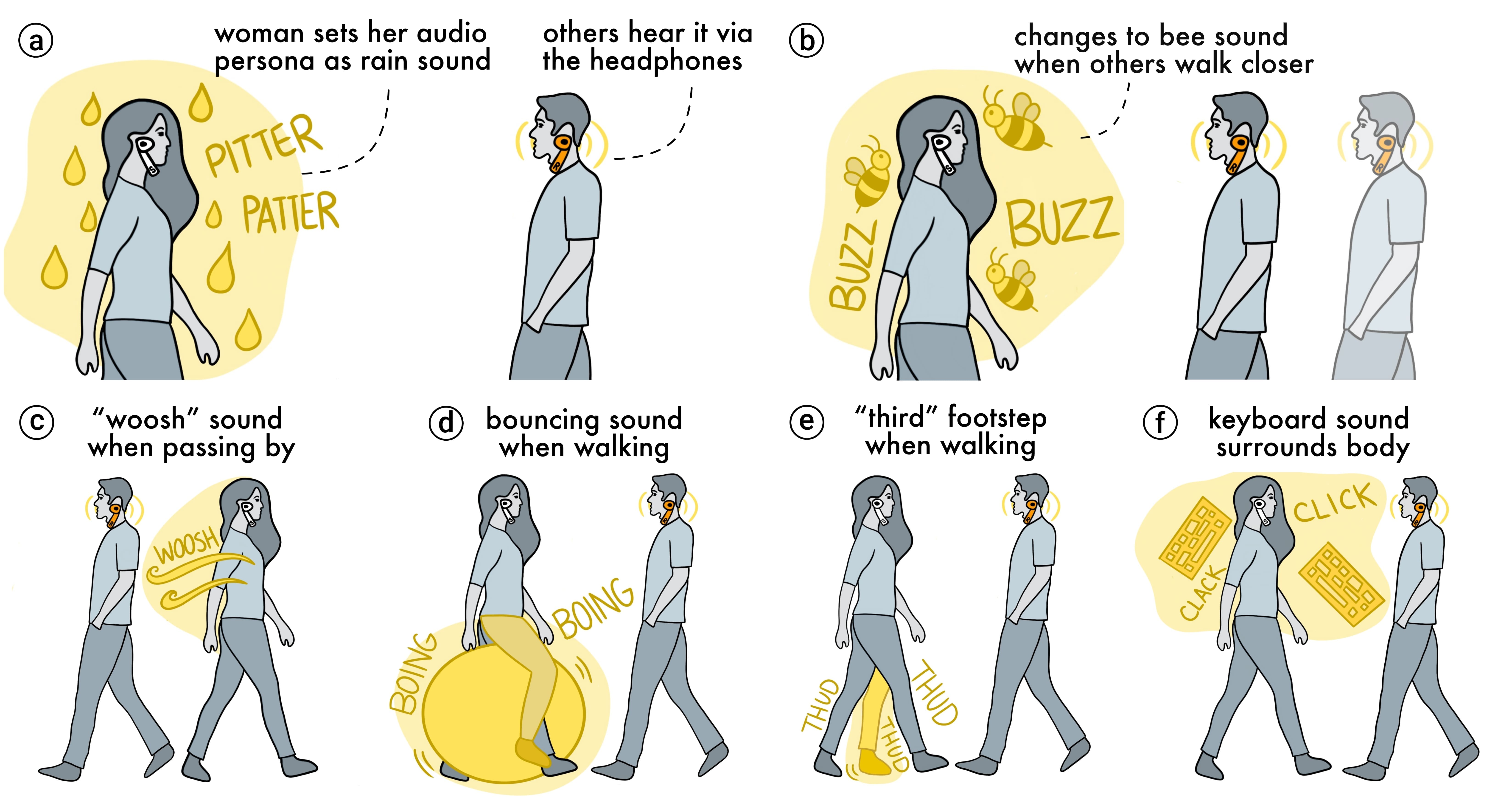}
  \vspace*{-20pt}
  \caption{Demonstrating the concept of audio personas, body-anchored audio cues to enhance face-to-face social interaction. Here, the yellow illustrations indicate the content of the sound and are not rendered visually. (a) The woman sets her audio persona as a rain sound surrounding the body and others can hear it through headphones. (b) As others walk closer, the woman changes the audio persona to a bee sound. Showcasing example designs of audio personas. (a) a "woosh" sound when passing by, (b) a bouncing sound when walking, (c) a third footstep sound when walking, and (d) a keyboard sound surrounds the body.}
  \Description{}
  \vspace*{-10pt}
  \label{fig:teaser}
  
\end{figure}

Despite growing efforts to enhance face-to-face interactions with digital cues, most explorations have concentrated on augmenting one’s visual appearance. However, our perception of others is inherently multisensory, involving not just how one looks, but also other sensory inputs such as their voice, footstep sounds, and their scent~\cite{fiore1993multisensory, fuller2007exploratory}. In this work, we explore the opportunities to enhance face-to-face interactions through the auditory channel. 

We introduce the concept of audio personas, body-anchored audio cues to augment social perception in face-to-face interactions. Here, the term \textit{persona} refers to the aspect of someone's character that is presented to others~\cite{jung2014two, goffman2002presentation}, which can vary depending on the social context~\cite{jung2014two, goffman2002presentation, leary1990impression, tajfel1974social}. Sound effects are commonly used in online games to provide immediate feedback for user actions like hitting a barrier, or collecting a reward, enhancing player engagement~\cite{liljedahl2011sound, lipscombc2013playing, kenwright2020there}. However, using audio in face-to-face social contexts is largely unexplored. In fact, psychological studies showed that exposure to emotion-evoking sounds can affect how people perceive others’ social attractiveness and emotions~\cite{may1980effects, hanser2015effects, logeswaran2009crossmodal}. For instance, women rated photos of men as more attractive when listening to positive music compared to when they heard negative music~\cite{may1980effects}. With the rise of spatial audio, the ability to spatially anchor sounds to individuals—similar to how visual filters personalize appearance—creates a novel means of personalized expression through sounds.

Figure~\ref{fig:teaser} visualizes a few examples of audio personas. For instance, one can choose buzzing bee sounds to indicate personal boundaries, or a "woosh" sound to mimic passing by like a breeze for fun and playfulness. One could also customize their footsteps to sound like bouncing balls or add a third footstep to express their love for gaming. Alternatively, keyboard-clicking sounds could be used to signal that the user is busy. To instantiate this concept, we first built a headphone-based prototype that utilizes multi-person tracking and multi-channel audio streaming. The user anchors sounds to their body and projects them to others in audio augmented reality.

We conducted two user studies to explore the concept of audio personas. Study \#1 (n = 64) examined whether audio personas are effective in influencing social perception, using a controlled, in-lab setup. Audio personas are designed to be body-anchored; therefore, as a between-subjects factor, we varied the spatial anchoring of sounds to assess the impact of this design choice. Participants interacted with two research assistants while hearing sounds either anchored to the assistants (i.e., audio personas) or to objects in the environment. As a within-subjects factor, we varied the valence of sounds to investigate how the emotional tone of audio personas influenced social perception. Overall, we found that participants were more likely to incorporate audio personas, body-anchored audio cues, into their impressions of others compared to the object-anchored audio. Moreover, participants perceived the person with a positive audio persona as more socially attractive, likable and less threatening than the person with a negative audio persona.

As Study \#1 demonstrated that audio personas can influence social perception, Study \#2 aimed to gain an understanding of use cases and design possibilities for audio personas. Through a design study with 8 audio designers, we identified intended use cases of audio personas in public and semi-public-private spaces as a way to manage social impressions (e.g., personality) and signal current states (e.g., emotions). We also elicited the creations of 48 audio personas from the designer participants and identified diverse design patterns in localization, triggers, dynamics and content of audio personas.

\vspace*{5pt}
Our key contributions are: 
\begin{itemize}
  \item We introduce audio personas, body-anchored audio cues to enhance social perception in face-to-face interactions. The concept opens up possibilities for engaging dynamic, multisensory cues to shape social perception.

  \item In a preregistered in-lab study (n = 64), we showed that audio personas can influence how one forms impressions of others. These impressions were shaped in alignment with the valence of the audio persona.

  \item Through a design study with audio designers (n = 8), we explored the social intentions and contexts for audio persona usage. This study also uncovered design possibilities for audio personas and identified key design patterns.

  \item We discussed insights from the user studies and laid out design considerations and directions for future explorations of audio personas. 

\end{itemize}

\section{Related Work}
Our work builds on prior research that uses digital media to enhance social perception, as well as ongoing efforts to integrate dynamic digital elements into face-to-face interactions. We extend previous research that employs auditory cues to support physical interaction, i.e., audio augmented reality (audio AR).

\subsection{Social Perception in Digital Media}
In everyday life, people choose clothing, apply makeup, and use perfume to influence how others perceive them. This process, often referred to as impression management~\cite{leary1990impression, goffman2002presentation}, plays a crucial role in shaping social interactions. By selecting their attire and accessories, individuals convey different aspects of their self-identity ~\cite{lennon1989clothing, branstetter1975see, miller1997dress, workman1991role, croijmans2021role}. 

Just as people carefully manage their social impressions in face-to-face interactions, they also do so in the digital world. On social media apps like Instagram and Snapchat~\cite{choi2018instagram}, individuals craft their online presences via selfies and profile photos~\cite{segalin2017your, siibak2009constructing}. Using digital filters and editing tools, users can easily enhance and transform appearances by smoothing skin, changing hair color, and adding digital accessories or makeup~\cite{siddiqui2021social, javornik2022lies, hong2020you}. Similarly, these digital filters have also been adopted in videoconferencing to augment remote communications~\cite{li2023filters, leong2021exploring}. 

In virtual worlds, such as those found in games and virtual reality (VR) environments, users can create and represent themselves as avatars. On platforms such as ~\textit{Second Life}~\cite{secondLife}, self-avatars serving as users' digital personas within the virtual space might or might not align with their physical social identity~\cite{martey2011performing, schultze2009avatar}. Game designers also map sound effects to player movements to indicate the outcomes of their actions~\cite{liljedahl2011sound, lipscombc2013playing, kenwright2020there}. These sounds provide immediate feedback, such as the noise of a sword hitting an enemy or the sound of footsteps, to enhance immersion. In first-person games such as ~\textit{Half-Life}~\cite{halfLife}, the introduction of avatar sound effects was found to increase gamer experience~\cite{nacke2010more}.

With the emergence of social VR platforms, such as ~\textit{Horizon}~\cite{horizon} and ~\textit{VR Chat}~\cite{vrchat}, users are able to directly embody the digital avatars that represent them~\cite{kilteni2012sense}. These avatars can not only reflect their desired appearance~\cite{kim2023or} but can also transcend their physical presence~\cite{yee2007proteus, bailenson2004transformed}. For instance, users can become giants~\cite{abtahi2019m}, have novel body parts~\cite{won2015homuncular} or even adopt non-humanoid avatars~\cite{krekhov2019illusion, ahn2016experiencing}. There is also a growing attention to how audio in VR, such as changes to the avatar's voice~\cite{choi2023exploring, kao2022audio, byeon2023avocus,Cheng2024Hearing} or the presence of footstep sounds~\cite{kern2020audio} impact immersive experience. Similarly, audio is gaining attention in remote collaboration contexts~\cite{yang2020effects, takayama2010throwing}. Researchers found that collaborating with avatars with self-similar voices led to higher ratings on the characters' likability and believability~\cite{guo2024collaborating}.

\vspace{-20pt}
~\subsection{Introducing Dynamic Digital Elements to Face-to-Face Social Interaction}
While digital media provides extensive flexibility in managing online social perception, the physical adornments we use every day are often static and do not adapt to evolving social contexts. Thus, researchers explored to introduce dynamic digital characteristics into physical interactions. 

Drawing inspiration from the filters used in social media and online meeting platforms~\cite{Rios2018}, researchers explored the application of visual filters in everyday augmented reality settings~\cite{chakrabarty2024exploring}. For example, by attaching AR markers onto clothing, users can see animated or interactive elements superimposed onto their physical attire~\cite{hakkila2017clothing, Roinesalo2016, kp2023weaving}. Additionally, mobile AR technology enables the digital modification of the appearance of physical jewelry and clothing, allowing users to experiment with different colors, patterns, and designs in real-time~\cite{Rantala2018, Roinesalo2016, Mackey2020, Fuste2018}. 

From a different angle, researchers also explored embedding wearable displays onto physical clothing to make physical adornments interactive and adaptive. An early prototype, BubbleBadge~\cite{falk1999bubblebadge}, was designed as a coat badge and was able to provide various information about the wearer. Similarly, accessories such as necklaces~\cite{buruk2021snowflakes}, handbags~\cite{colley2016smart}, and rings~\cite{buruk2021snowflakes} can also be embedded with LEDs and small displays to change appearances dynamically. Researchers also explored technical possibilities of making it battery-free~\cite{dierk2018alterwear}, interfacing directly with the skin~\cite{Wang2017}, and integrating with existing textile printing methods~\cite{howell2016biosignals, kan2015social, devendorf2016don}.

While our perception of others involve both visual (e.g., facial expressions, body language, clothing) and non-visual cues (e.g., voice, motion sounds, scents)~\cite{fiore1993multisensory, fuller2007exploratory}, prior works have primarily focused on augmenting visual appearance. Our work broadens the range of modalities used to enhance physical social interactions with digital cues.

\vspace{-15pt}
~\subsection{Dynamic Auditory Cues to Influence Social Perception}
In this work, we propose using audio cues, anchored to the body, to augment social perception in face-to-face interactions. This approach is inspired by previous research in audio augmented reality~\cite{yang2022audio}, which demonstrated that altering environmental and body-generated sounds can affect how individuals perceive both their surroundings and bodies.

The term "soundscape" refers to an acoustic environment perceived by human ears~\cite{pijanowski2011soundscape}. In audio augmented reality, soundscapes were used to deliver location-based sonic experiences to enhance the understanding of new sites~\cite{rozier2000here, mcgookin2012pulse}, support wayfinding~\cite{mcgookin2009audio} and facilitate interactions with everyday objects~\cite{yang2019hearing}. Sikora et al. found that visitors exposed to an ancient medieval soundscape (e.g., horse-riding sounds) at an archaeological site showed higher arousal than those without the soundscape.~\cite{sikora2018soundscape}. By introducing audio effects such as delay and frequency changes to environmental sounds ~\cite{storek2015modifications, watanabe2020manipulatable}, researchers developed systems to "warp" the perception of the surrounding space. 

Our bodies generate sounds both within themselves (e.g., breathing, heartbeat) and through interactions with the environment (e.g., tapping sounds). Tajadura-Jiménez et al. introduced the concept of "sonic self-avatar," emphasizing that self-representations in mixed reality environments should incorporate body-centered sounds~\cite{tajadura2018principles, tajadura2008auditory}. When body actions are sonified, the sonification can influence users' mental representations of their bodies~\cite{tajadura2015action, d2024soniweight, tonetto2014modifying, frid2016interactive}, perception of objects~\cite{furfaro2015sonification}, and even gaits and motion~\cite{tajadura2015light}. For example, amplifying the high frequencies of footstep sound can create the self-perception of having a thinner body~\cite{tajadura2015light} and sonifying body movement with descending pitch increases movement acceleration ~\cite{singh2024pushed}. These audio-based sensory feedback devices were further explored in the context of physical activity and rehabilitation~\cite{ley2024co, turmo2024body, gomez2020enriching}. 

Building on previous research that showed modifying environmental and body-centered sounds can enhance perceptions of space and self, this work taps into the exploration of audio in social contexts. Previous psychology research has shown that emotion-evoking music can influence how others are perceived in photographs, affecting factors like social attractiveness~\cite{may1980effects, hanser2015effects}, perceived emotional states~\cite{logeswaran2009crossmodal, ignacio2021music, carroll2005priming} and even interpersonal distance~\cite{tajadura2011space}. In computer-mediated communication, audio serves as a form of social signaling. Previous research explored music-sharing systems~\cite{golan2018icebreakware, mueller2002transparent, gurion2013audio}, sonic identification to aid in remembering others~\cite{choi2011sound, choi2017designing} and background auditory cues linked to people's activities~\cite{mynatt1998designing}. However, none of the previous work has explored the use of body-anchored sounds to facilitate dynamic social expressions in face-to-face interactions, opening up a wide range of design opportunities beyond what has been explored so far.

\section{Proof-of-concept System}
\begin{figure}[hbt!]
  \includegraphics[width=\linewidth]{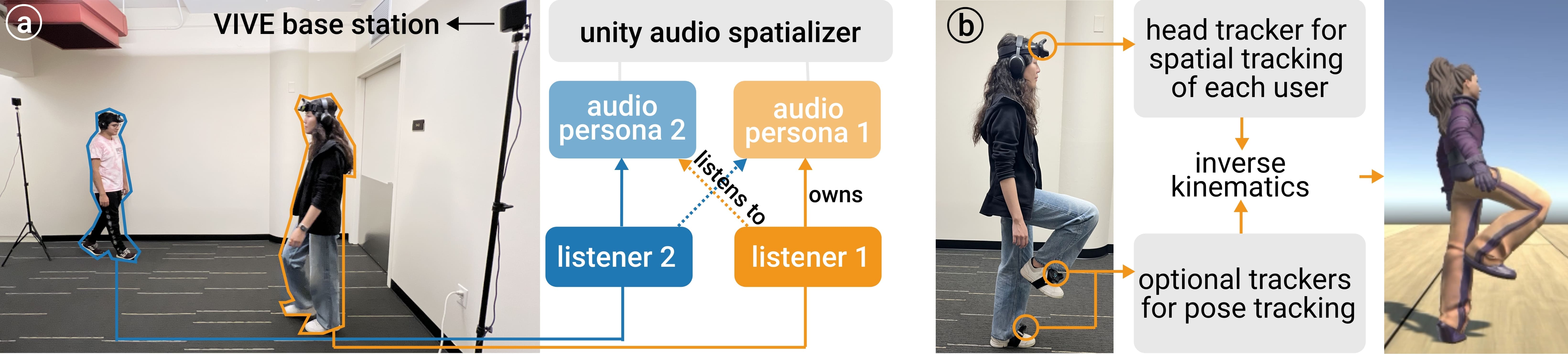}
  \vspace{-15pt}
  \caption{Overview of the proof-of-concept system. (a) Spatial tracking using ~\textit{VIVE} base stations. Here, each user wears a ~\textit{VIVE} tracker on the forehead, which is used to track their spatial location in the space. Each user in the system acts as a receiver and a sender for the audio persona. All audios are rendered based on the spatial location of each user using ~\textit{Unity} engine. (b) Besides the head tracker, users can optionally wear trackers on other parts of their body, such as the feet. These trackers help capture the user's fine-grained body motions and trigger motion-specific audio personas, such as changing the sound of their footsteps.}
  
  \Description{}
  
  \label{fig:system}
\end{figure}

To deliver the audio persona experience, we implemented a proof-of-concept system that supports: (1) spatial and motion tracking of each user and (2) multi-channel audio streaming. The setup enables multi-user interactions, allowing users to both project and listen to audio personas. Figure~\ref{fig:system} shows the system setup. 

\subsection{Spatial and Motion Tracking}
We used \textit{HTC VIVE} base stations and trackers 2.0~\cite{HTC} for spatial and motion tracking. Each user wears a \textit{VIVE} tracker on their forehead to approximate body location and orientation. Optionally, additional trackers could be attached to body joints (e.g., feet, hands) for limb tracking. Each tracker records 6DOF tracking data, including position ($x, y, z$) and orientation ($roll, pitch, yaw$). Tracking data was processed in real-time in \textit{Unity} using the inverse kinematics pipeline, \textit{Final IK}\cite{finalik}, to create a digital twin of the user's physical movements, as shown in Figure~\ref{fig:system}b. The audio persona was rendered by anchoring the audio source to a specific body joint of the user avatar (e.g., head, feet) within ~\textit{Unity}. This setup supports customizable audio persona experiences, regardless of the number of trackers used.

Tracking the user's head alone offers a wide range of interaction possibilities. Given the head locations of each user ($x_{h1}, x_{h2}, x_{h3}...$), the system enables proximity-based interactions by calculating the distances between head trackers ($d_{12} = ||x_{h1} - x_{h2}||$,  $d_{13} = ||x_{h1} - x_{h3}||...$). The system can then trigger audio personas, such as a bee sound, when users are at a certain distance. Moreover, by computing the change in head position (\(\mathbf{x}_{hi}(t)\)) and orientation \(\mathbf{R}_{hi}(t)\) over time, the system can derive the speed of the user's movement (e.g., \(\mathbf{v}_{hi}(t) = \frac{d\mathbf{x}_{hi}(t)}{dt}\)). This allows for compound effects: for instance, users can project a "woosh" sound when they move fast and approach another user.

With additional body trackers, more fine-grained, movement-based audio personas can be implemented. For example, a footstep-based audio persona (e.g., giant stomps) involves assigning virtual colliders to the avatar's feet ($f_1, f_2$) and the virtual floor ($z_{floor}$), pre-calibrated to match the actual floor height in ~\textit{Unity}. Collisions are detected when the vertical position of a foot ($z_{f_i}$): $z_{fi} + \sigma \le z_{floor}$, where $\sigma$ is collision detection threshold.

\subsection{Audio Rendering and Multi-channel Audio Streaming}
We further used ~\textit{Unity} as the audio engine for the system, which provides control over 3D sound dynamics (e.g., min/max distance and volume rolloff). For realistic spatial audio rendering, we used the ~\textit{Meta XR Spatializer}~\cite{metaaudio}.

To support multi-user audio persona experience, we implemented multi-channel audio streaming. This means that each user using the system is a sender, broadcasting their audio persona to others while simultaneously hearing everyone else's audio personas. To accomplish this, each user avatar in ~\textit{Unity} was attached with one (or more) audio sources and one audio listener. 

Specifically, we extended ~\textit{Unity}'s default capabilities by implementing support for multiple virtual listeners, going beyond the standard single-listener setup. We built on top of the plugin developed by Gael Vanhalst~\cite{MultiAudioListener}, which provided a foundation for this capability. Additionally, we integrated the ~\textit{Audio Stream} pipeline~\cite{AudioStream}, which supports PCM audio streaming, to further route the audio output to separate audio devices connected to the host computer. Each audio persona the user hears is spatialized based on the location of the person who set that specific audio persona. In the current setup, the maximum number of users the system can support is limited by the number of output devices the host computer can connect to.

\section{Study \#1: Can Audio Personas Affect Social Perception?}
To explore whether audio personas are effective in influencing social perception, we conducted a preregistered study to examine key design components of audio personas—specifically audio anchoring and audio valence—in a controlled setting. 64 participants were invited to the lab to take part in an icebreaker activity with two research assistants.

\subsection{Condition Design}
A mixed-factorial study design was adopted in the study, with one between-group and one within-group factor. The supplemental video illustrates the auditory experience of each condition.

\subsubsection{\textbf{Between-group: audio anchoring.}} Audio personas are proposed to be body-anchored. Thus, to understand the effect of audio personas on social perception, it's essential to understand how this spatial anchoring plays a role.

\begin{figure}[hbt!]
  \includegraphics[width=\linewidth]{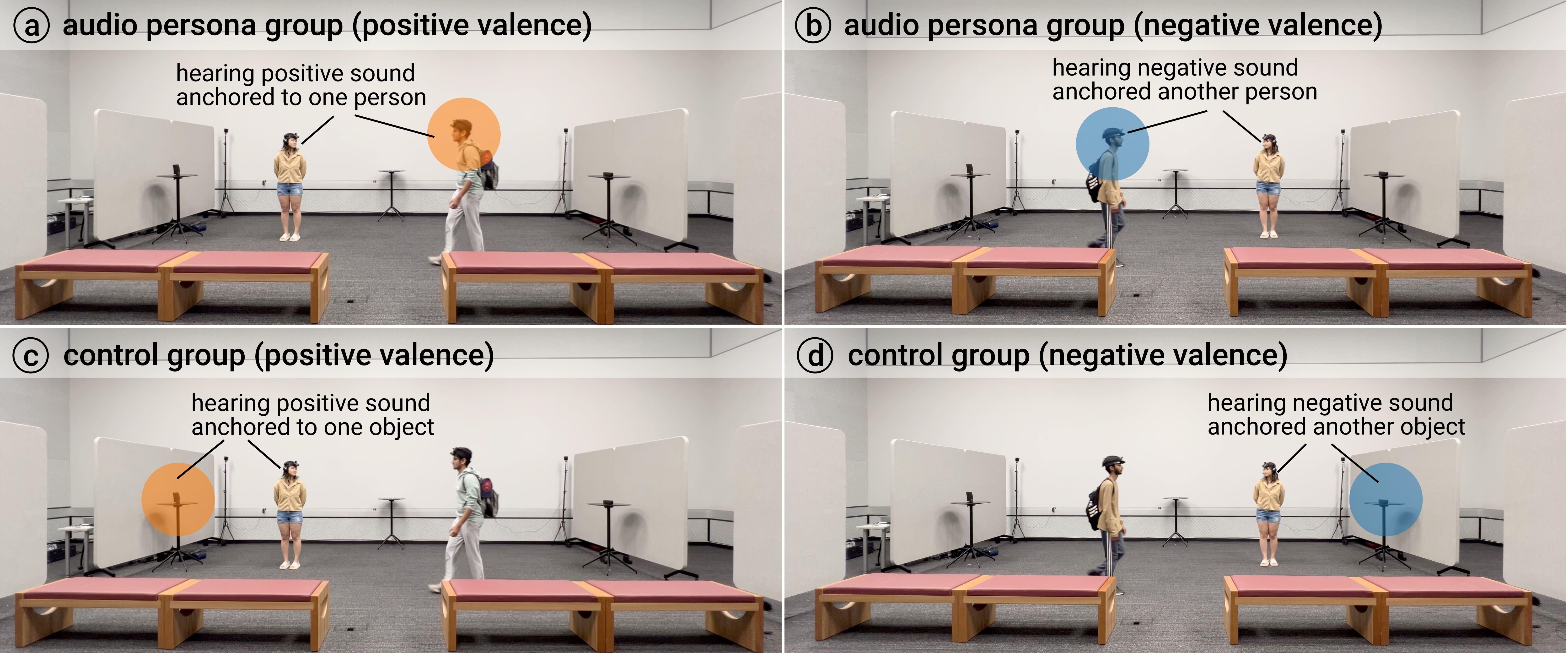}
  \caption{Condition design in Study \#1. (a) and (b) show the experience of the ~\textit{audio persona} group, where the participants heard sounds anchored to the research assistant they interacted with. One research assistant had a positive audio persona, while the other had a negative audio persona. (c) and (d) show the experience of the ~\textit{control} group, where participants heard audio anchored to objects in the space (i.e., speakers). One speaker had a positive sound anchored to it, and the other had a negative sound anchored to it. In both groups, the research assistants entered the space from different sides of the room, and each walked back and forth during the task. Participants always stood at the opposite side of the room from where the research assistant they would observe first entered.}
  \Description{}
  \label{fig:study2_conditions}
  \vspace{-10pt}
\end{figure}

Participants were divided into two groups. Half of the participants were assigned to the ~\textit{audio persona} group (Figure~\ref{fig:study2_conditions}a, ~\ref{fig:study2_conditions}b), in which participants heard sounds anchored to each research assistant. As the research assistant walked past participants (e.g., moving from left to right or vice versa), their audio persona shifted in locations corresponding to their movements.

In the ~\textit{control} group, participants interacted with research assistants without hearing their audio personas. Instead, participants heard sounds anchored to objects (i.e., speakers) in the space. We chose to include sounds in the ~\textit{control} group as previous research showed that the mere presence of audio can already influence social emotional perception and behavior~\cite{tajadura2011space, logeswaran2009crossmodal, carroll2005priming}. This control design thus better examines the exact effect of "body-anchored" audio. Anchoring sound to objects is also commonly used in audio augmented reality~\cite{mcgookin2009audio, yang2019hearing, sikora2018soundscape} and the auditory experience in this group is similar to hearing sounds played from a physical speaker. The research assistants' movements would not affect the sound.

\subsubsection{\textbf{Within-group: audio valence.}}
Prior research has shown that audio valence can evoke distinct emotions in listeners~\cite{zentner2008emotions, juslin2008emotional, tajadura2008auditory}. Therefore, Study \#1 also examined how the emotional component of the sounds used in audio personas influences social perception. Specifically, participants heard different types of sounds when interacting with each of the two research assistants. In the ~\textit{audio persona} group, one research assistant had a \textit{positive audio persona}, and the other had a \textit{negative audio persona}. 

In the \textit{control} group, one object had a positive-valence sound (\textit{positive control}), and the other had a negative-valence sound (\textit{negative control}). Participants were instructed to stand on the opposite side from where the research assistants entered, ensuring they heard one sound while observing one assistant and a different sound while observing the other (Figure~\ref{fig:study2_conditions}). This design was also used for the ~\textit{audio persona} group to ensure consistency.

~\textit{\textbf{Nested factor: audio content.}} To ensure generalizability of the results, we incorporated diverse sounds. Specifically, we included two different sounds for each valence condition (positive and negative), resulting in a collection of two positive and two negative sounds. For each trial, one positive and one negative sound were used.

We chose audio content from established audio databases to ensure control over the valence of sounds. Specifically, we used IADS-2 databse~\cite{bradley2007international} and its expanded version (IADS-E)~\cite{yang2018affective}, following prior work~\cite{viinikainen2012representation, masullo2021questionnaire}. We focused on nature and animal sounds due to their universal familiarity, and sounds were selected based on valence ratings provided in the database. The positive-valence sounds selected for the study were brooks (valence: 7.28) and forest (valence: 6.91), and negative-valence sounds selected were buzzing (valence: 1.92) and hurricane (valence: 2.18).

\subsubsection{\textbf{Counterbalancing}}
We counterbalanced the audio valence (positive vs. negative) and the audio content (two options per valence) that participants heard during interactions with each research assistant. Additionally, we counterbalanced the side of the room from which each research assistant entered, which determined the participant’s standing position (as described in Section 5.1.2). Therefore, the sample size needed to be a multiple of $2$ (audio valence) × $2$ (audio content per valence) × $2$ (entry sides) × $2$ (research assistants) × $2$ (audio anchoring) = 32, to account for all combinations across both groups. Additionally, we randomized the order in which each research assistant entered and the side of the room where each speaker was located.

\subsection{Other Study Design Considerations}

\subsubsection{\textbf{Study instruction}} We introduced the study as an exploration of audio augmented reality for social interaction. Participants were informed that they would interact with two researchers and might hear sounds people chose to anchor to themselves or sounds anchored to objects in the space. The purpose of these sounds (e.g., signaling personality or emotions) was not disclosed to avoid biasing participants' reactions with preconceived expectations, in line with prior work on the psychological impact of multisensory cues~\cite{zhao2023affective, costa2016emotioncheck}.

\subsubsection{\textbf{Task design}} Participants and two research assistants took part in an activity called "Something’s Different"~\cite{krueger2009ice}. We adapted this common icebreaker task to foster spontaneous interactions. In the activity, participants needed to identify a change (e.g., clothing or accessories) in each research assistant between two sessions. In each session, research assistants entered the room individually, walking back and forth in front of the participants. The changes to each research assistant were always on their name tags to remain consistent across participants: one switched from a yellow name tag to blue, and the other from a blue name tag to yellow.

\subsubsection{\textbf{Familiarization phase}} 
Prior to the icebreaker, all participants completed a familiarization phase. This was intended to get all participants comfortable wearing the system, and also to prevent the control group from mistakenly attributing object-anchored sounds to the person. During this phase, participants were asked to walk around the space and identify whether any sounds were present and where they originated. The sounds were configured exactly as they would be during the icebreaker, and the control group heard sounds as they approached the corresponding objects. The research assistant was not present during this phase to ensure that participants encountered the research assistant for the first time during the main session. As a result, the audio persona group did not hear any sounds during this phase. All participants were able to identify whether sounds were present and, if so, determine their sources before proceeding with the icebreaker.

\subsubsection{\textbf{Research assistants}} All participants interacted with the same two male, Asian research assistants, who wore identical clothing across study sessions to minimize the influence of attire on social perception. We chose to prescript the research assistants' behavior in the activity to minimize confounding factors and isolate the specific impact of the audio personas on social perception. Both research assistants were trained to maintain neutral body language and facial expressions in the study, similar to~\cite{tajadura2011space}. They were instructed to walk at a steady pace, avoid expressive gestures and facial expressions, and maintain forward-facing eye gaze. An experimenter provided real-time feedback during repeated practice to ensure consistent and neutral behavior. In the study sessions, research assistants were blinded about the conditions assigned to the participants.

\subsection{Apparatus}
\begin{figure}[hbt!]
  \includegraphics[width=\linewidth]{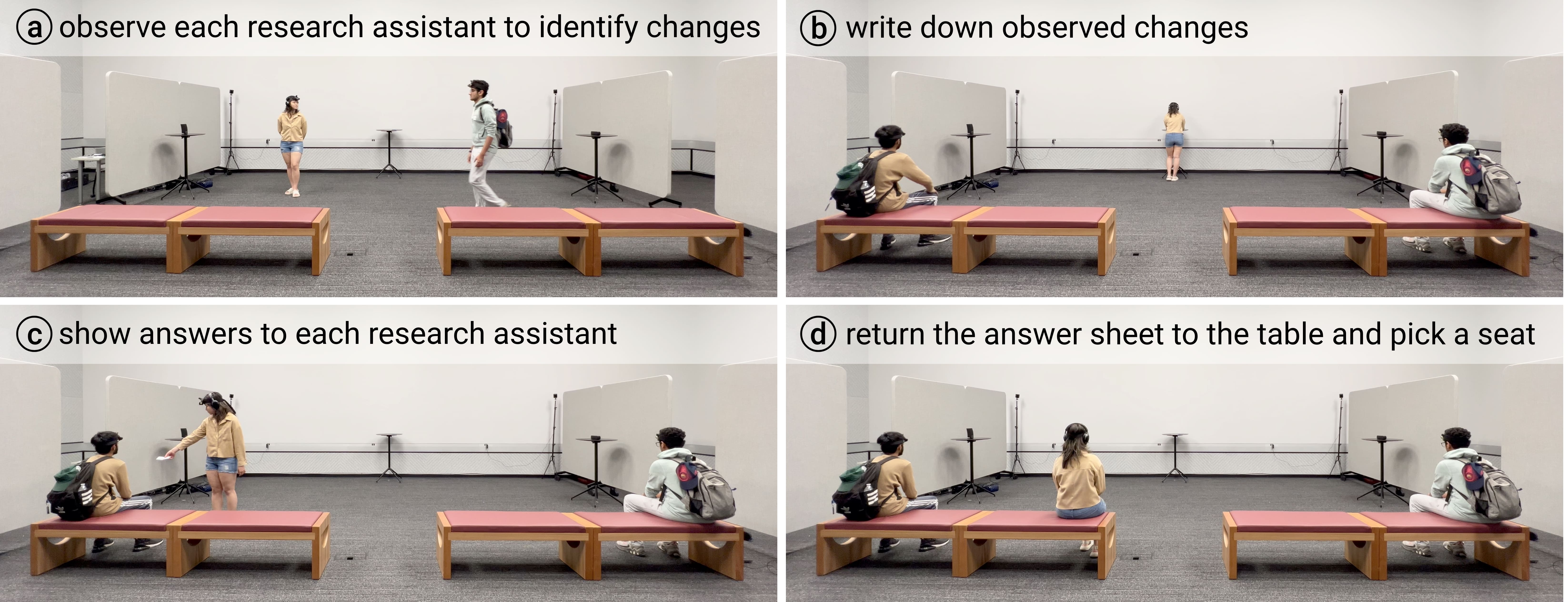}
  \caption{Key procedures in the Study \#1: (a) The participant observed each research assistant twice as they walked back and forth, to identify a change in each of them. (b) The participant wrote down the observed changes using the materials provided at the table in the back of the room. (c) The participant presented their answers to each research assistant. (d) The participant returned the answer sheet to the table and chose a seat to sit.}
  \Description{}
  \label{fig:study2_procedure}
\end{figure}

Figure~\ref{fig:study2_procedure} shows the study setup. The study space was 7 meters (length) x 5 meters (width) and was tracked by four \textit{HTC VIVE base station 2.0} units. Participants and research assistants all wore a head-mounted \textit{HTC 2.0 VIVE tracker} for 6DOF head tracking at 80Hz. Two standing locations were marked on the floor, 2.4 meters apart. At each standing location, participants were positioned 1.5 meters from the audio source anchored to the nearest object. To maintain social comfort~\cite{hall1966hidden}, the study space was arranged so that the minimum distance between participants and research assistants passing by was also approximately 1.5 meters.

Both body-anchored and object-anchored sounds were configured with a linear roll-off, with a minimum distance of 0.5 meters and a maximum distance of 2 meters. A 2-meter trigger distance was chosen following E.T. Hall’s "social distance" range~\cite{hall1963system}. Specifically, the audio persona group began hearing sound from a distance of 2 meters, with the volume increasing as the research assistants moved closer to them. The control group participants consistently heard the audio anchored to objects, regardless of the location of the research assistants. Details about sound characteristics are provided in the supplemental materials. Participants wore \textit{Bose QuietComfort Ultra Headphones} in \textit{aware} mode, allowing them to hear ambient sounds from the environment.

\subsection{Procedure}
Upon arriving at the lab, participants completed a consent form. Then, the experimenter introduced both research assistants to the participants by name and calibrated everyone’s head trackers based on their height. The participants were then introduced to the icebreaker task "Something’s Different", where their goal was to identify the change in each of the research assistants between two observation sessions. Before starting the task, researchers asked participants to walk around the study space to become familiar with it. Additionally, researchers showed participants the designated standing locations for observing the research assistants, as described previously.

The task included two observation sessions. In each session, research assistants took turns walking back and forth in front of the participants five times. Only one research assistant was present in the room at a time. After both sessions, participants were instructed to turn to the table behind them and use the provided pen and paper to write down the changes they observed in each research assistant. Meanwhile, both research assistants entered the room again, and each sat on a bench in the space, as shown in Figure~\ref{fig:study2_procedure}.

After writing their answers, participants were instructed to walk to each research assistant and show them the results. The research assistant responded with a nod after reading. Once participants had shown their answers to both research assistants, they were asked to place their answer sheets back on the table and return to sit on one of the benches. During this phase, behavioral measures including participants’ physical distance from each research assistant and their seat selection were recorded. Participants could still hear the sounds anchored to the person or object if they were within the trigger distance. At the end, experimenters entered to reveal the correct answer to the change and guided the participants to another room to complete a post-task questionnaire.

The study was approved by Stanford Institutional Review Board (IRB \# 71992). The full study took approximately 30 minutes, and all participants were compensated \$15 for their time.

\begin{table}[]
\begin{tabular}{@{}ll@{}}
\toprule
\textbf{Construct} & \textbf{Question} \\ \midrule
\multirow{3}{*}{\textbf{likability} $(r_{\alpha} = 0.75)$} & I think this person is competent. \\
 & I think this person is likable. \\
 & I think this person is trustworthy. \\ \midrule
\multirow{2}{*}{\textbf{perceived threat $(r_{SB} = 0.6)$}} & I think this person is dominant. \\
 & I think this person is threatening. \\ \midrule
\multirow{4}{*}{\textbf{interpersonal attraction} $(r_{\alpha} = 0.91)$} & I would enjoy completing a task with the person. \\
 & I would like to interact with the person again. \\
 & I would like this person. \\
 & I would get along with this person. \\ \midrule
\multirow{5}{*}{\textbf{perceived positive emotions} $(r_{\alpha} = 0.81)$} & I think the person feels alert. \\
 & I think the person feels determined. \\
 & I think the person feels inspired. \\
 & I think the person feels attentive. \\
 & I think the person feels active \\ \midrule
\multirow{5}{*}{\textbf{perceived negative emotions} $(r_{\alpha} = 0.72)$} & I think the person feels hostile. \\
 & I think the person feels ashamed. \\
 & I think the person feels upset. \\
 & I think the person feels afraid. \\
 & I think the person feels nervous. \\ \bottomrule
\end{tabular}
\vspace{5pt}
\caption{Self-report likert scale questions in the post-study questionnaire. We report reliability scores for constructs with more than two items using Cronbach's alpha ($r_{\alpha}$) and two-item constructs using Spearman-Brown ($r_{SB}$).} 
\label{table:table1}
\vspace{-20px}
\end{table}

\subsection{Measures}

We adopted both self-report and behavioral measures to understand how audio personas might affect social perception. 

\subsubsection{\textbf{Self-report measures}}
We started with \textbf{impression drawings} and \textbf{impression writings}. Drawing is a common technique in psychology to measure social perception~\cite{kamphaus1991draw, cherney2006children}, and we used drawing to understand if participants formed impressions of others with explicit consideration of their audio persona. This was contrasted by the drawings from the control group, where the sounds existed but were not spatially associated with the person. In support of the drawings, participants also wrote down their impressions of each research assistant using at least five sentences, referred to as impression writings. 

To capture the nuances of social impressions beyond the incorporation of sounds, we included constructs such as likability judgments, threat perception, and interpersonal attraction, inspired by prior work~\cite{fauville2022impression}. All constructs were rated on a Likert scale from 1 (not at all) to 5 (very much), shown in Table~\ref{table:table1}. ~\textbf{Likability judgments} included the dimensions of attractiveness, competence, extroversion, likability, trustworthiness~\cite{todorov2013validation, fauville2022impression}. \textbf{Threat perception} was measured by two dimensions, threat and dominance~\cite{todorov2013validation, fauville2022impression}.~\textbf{Interpersonal attraction} was measured by combining both social attraction (e.g., “I would get along with this person”) and task attraction (e.g., “I would like to interact with the person again.”), resulting in a 4-item construct~\cite{fauville2022impression, davis1979consequences}. 

Moreover, we examined whether audio personas influence how people interpret the emotional states of others, as prior work has shown that music can shape the perceived emotions of individuals depicted in photographs. ~\cite{logeswaran2009crossmodal, hanser2015effects}. Specifically, we used the I-PANAS-SF scale~\cite{karim2011international}, which includes 5 positive (active, attentive, alert, determined, inspired) and 5 negative terms (hostile, ashamed, upset, afraid, nervous), shown in Table~\ref{table:table1}.

\subsubsection{\textbf{Behavioral measures}}
Two behavioral measures were collected during the icebreaker activity: interpersonal distance and seat choice. \textbf{Interpersonal distance} was determined by calculating the Euclidean distance between the head positions (x and z axes) of the participant and each research assistant. Head positions and rotations were continuously tracked throughout the study session using head-mounted trackers. Specifically, we focused on the minimum interpersonal distance when participants were instructed to walk toward each research assistant to present their answer sheets. This task approximates the active approach paradigms used in prior work~\cite{tajadura2011space}. In addition, we recorded which bench participants chose to sit on at the end of the icebreaker (i.e., \textbf{seat choice}), with each research assistant sitting on one. The selection of a seat closer to one research assistant over the other reflects the degree of social closeness participants felt~\cite{hayduk1983personal}.

\subsubsection{\textbf{Overview of analysis}}
To analyze impression drawing and writings, two researchers independently coded the drawings and writings to analyze if participants referenced sounds through verbal or visual representations and resolved discrepancies through discussions. As exploratory measures, we analyzed the count of sound-related terms and sentiments in impression writings.

For self-report rating measures, we first assessed reliability using Cronbach's alpha for constructs with more than two items and Spearman-Brown formula for two-item constructs. Initial validation showed that reliability for likability (r$_{\alpha} = 0.62$) and threat perception (r$_{\alpha} = 0.60$) were lower than 0.7. A Principal Component Analysis identified two likability items (attractiveness, extroversion) loading in a different direction, which were dropped; this raised reliability to r$_{\alpha} = 0.75$. For the two-item threat perception construct, individual item results were reported, due to the inability to improve reliability via factor analysis. We then used mixed ANOVA with audio anchoring as a between-group factor and audio valence as a within-group factor to analyze each of the self-report rating measures.

For the behavioral measures, we also used mixed ANOVA to analyze the effect on minimum interpersonal distance. For the seat choice measure, since participants chose between one of the two benches to sit, we used a chi-square test to examine the relationship between audio valence and seat choice in each condition group.

\subsection{Hypotheses}

We preregistered the hypotheses through Open Science Foundation\footnote{https://osf.io/dzhca} and outlined the key hypotheses below. We rephrased the hypotheses from the language originally used in the preregistration to ensure clarity in this paper.

Audio personas are body-anchored audio cues, creating spatial alignment between the person and the sound. From a multisensory cue combination standpoint~\cite{spence2013just, zaki2013cue}, such spatial alignment increases the likelihood of associating the sound as relevant information of that person. Thus, when considering its potential impact on impression formation, we hypothesized that ~\textit{audio persona} group are more likely to incorporate visualizations of audio into their impression drawings of others, compared to ~\textit{control} group, who heard sounds anchored to the objects (\textbf{H1}). 

The valence of audio can trigger different emotional responses~\cite{zentner2008emotions, juslin2008emotional, tajadura2008auditory}. Thus, when looking into the valence of sound used in audio personas, we hypothesized that people would form a more positive impression of a person with a positive audio persona compared to a person with a negative audio persona (\textbf{H2}). Even though ~\textit{control} group didn't experience sound spatially anchored to the person, the temporal alignment introduced by hearing sounds while interacting with a person might already have an effect on social perception, based on prior work~\cite{tajadura2011space, logeswaran2009crossmodal, carroll2005priming}. Thus, we also hypothesized that people would form a more positive impression of others when there is a positive sound during their interactions, compared to when there is a negative sound, with the sounds anchored to objects in the environment (\textbf{H3}).

When comparing social impressions and perceived states between the two groups, we evaluate whether the spatial binding introduced by audio personas could enhance the effect of audio valence. We hypothesized a significant interaction between the audio anchoring and audio valence variables. Specifically, we hypothesized that a person with a negative sound anchored to them (i.e., negative audio persona) would create a more negative impression on others compared to when the negative sound is anchored to an object (\textbf{H4}). Similarly, we hypothesized that a person with a positive sound anchored to them (i.e., positive audio persona) would create a more positive impression on others compared to when the positive sound is anchored to an object (\textbf{H5}).

\subsection{Participants}
Following preregistration, we recruited 66 participants who were over 18 years old and had no hearing deficit. Two participants' data were excluded from the following analysis because of technical errors. Thus, the final sample size is 64 participants: 36 self-identified as females, 27 self-identified as males, and 1 self-identified as non-binary. The average age of participants was 24.7 years old ($SD = 6.3$). 14 participants identified as Caucasian, 3 identified as African American, 2 identified as Native American, 43 identified as Asian, 7 chose "other" and 2 chose "prefer not to say". The total added to more than the sample size as participants could select more than one ethnicity. None of the participants knew the research assistants before the study.

\subsection{Overview of Findings}
We organized the study findings into four sections. \textbf{Finding \#1, Inclusion of Audio in Impression Formation}, examines whether participants considered audio as part of their impression of others, manifested in the impression drawings and writings. \textbf{Finding \#2, Impact on Social Impression}, presents self-reported measures related to social impressions (social attraction, perceived threat, likability judgments), and sentiment analysis of the impression writings. \textbf{Finding \#3, Impact on Perceived Emotional States}, includes participants’ ratings on the perceived emotional states of others. \textbf{Finding \#4, Impact on Social Behavior}, reports participants’ seat choice selection and interpersonal distance.

\subsection{Finding \#1: Inclusion of Audio in Impression Formation}

We found that ~\textbf{20 out of 32 (62.5\%)} participants in ~\textbf{~\textit{audio persona} group} explicitly described the sounds they heard during interactions with others in their impression drawings or writings. Of these, 11 visualized the sounds in impression drawings, 18 mentioned them in impression writings, and 9 included descriptions of sounds in both.

In contrast, only ~\textbf{6 out of 32 (18.75\%)} participants in ~\textbf{~\textit{control} group} explicitly described the sounds they heard during interactions with others in their impression drawings or writings. Of these, 2 visualized the sounds in impression drawings, 6 mentioned them in impression writings, and 2 included descriptions of sounds in both.

\begin{figure}[hbt!]
  \includegraphics[width=\linewidth]{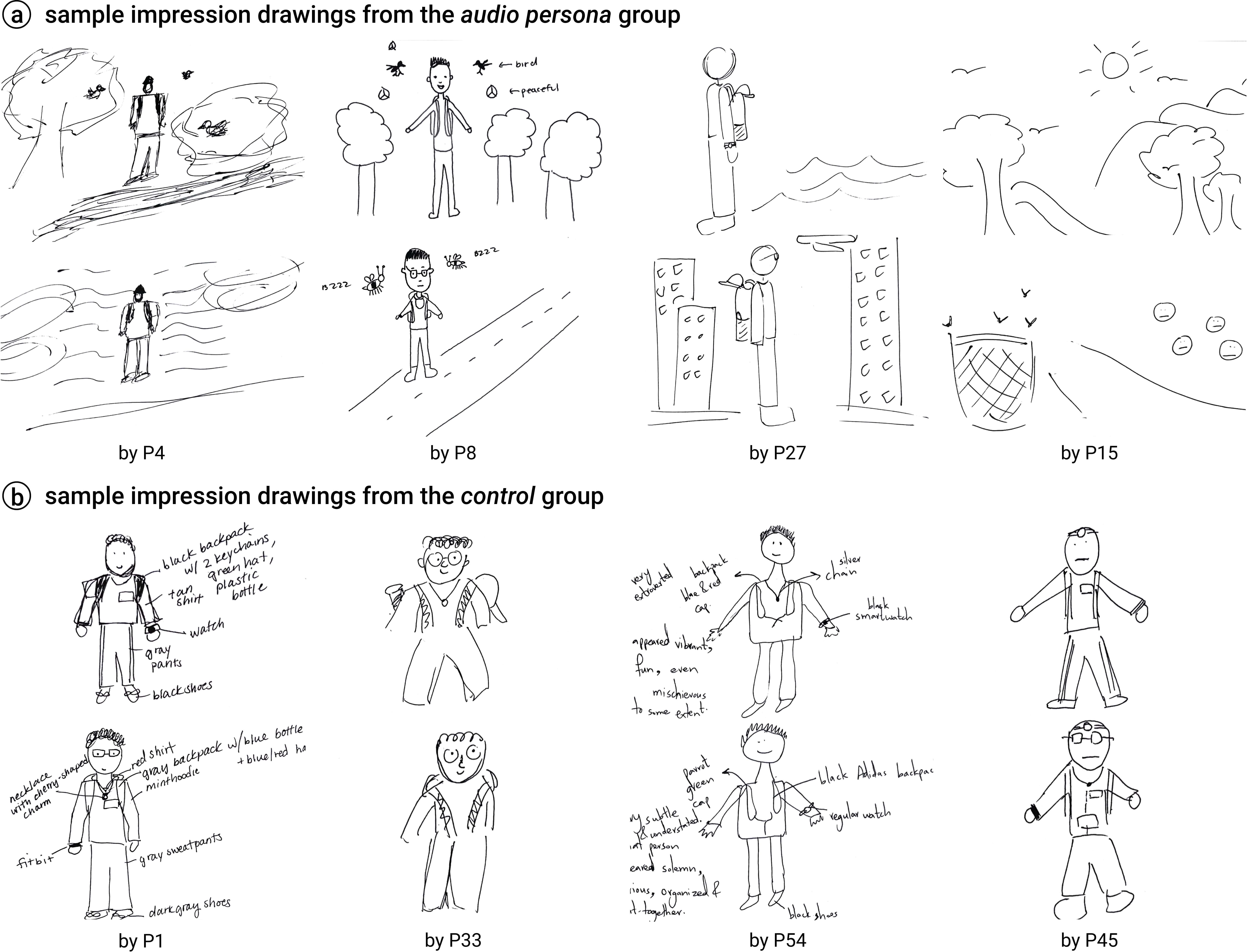}
  \caption{A selection of impression drawings from participants in the ~\textit{audio persona} and ~\textit{control} group. Participants in the ~\textit{audio persona} group tended to incorporate sounds they heard as part of their impression drawings, while participants in the ~\textit{control} group focused mostly on visual appearances. Each column represents drawings by the same participant. In (a) and (b), the top row shows impression of the research assistant paired with a positive sound, and the bottom row shows impression of the research assistant that was paired with a negative sound. Names of the research assistants were redacted from the drawings.}
  \Description{}
  \label{fig:impression_all}
\end{figure}

We applied a Z-test to compare the proportions of participants that include sounds in their impression drawings or writings of others. We found that the ~\textit{audio persona} group were significantly more likely to include visualizations of sounds in the impression drawings compared to the ~\textit{control} group ($p = 0.005$). Similarly, ~\textit{audio persona} group were significantly more likely to describe sounds in the impression writings than ~\textit{control} group ($p = 0.002$). These findings confirm ~\textbf{H1}. 

Figure~\ref{fig:impression_all} shows a selection of impressions drawings made by participants in both groups. In \textit{audio persona} group, participants often visualized the content of the sounds they heard, such as "forest" or "water" associated with positive audio personas, and "bees" or "hurricane" for negative audio personas. In contrast, in the ~\textit{control} group, when the sounds were anchored to objects, participants tended to focus on drawing the person's visual appearance.

\begin{table}[hbt!]
\begin{tabular}{@{}lcc@{}}
\multicolumn{3}{l}{\textbf{Word count of "sound", "sounds" and "audio" in impression writings}} \\ \midrule
                                 & control group              & audio persona group              \\ \midrule
with positive sound              & 4                          & 23                              \\
with negative sound                   & 5                          & 24                              \\
total                            & 9                          & 47                              \\ \bottomrule
\end{tabular}
\vspace{5pt}
\caption{The number of times the words "sound," "sounds," and "audio" were used in the impression writings.} 
\vspace{-20pt}
\label{table:table2}
\end{table}

As an exploratory measure, we analyzed the frequency of the words "sound," "sounds," or "audio" in participants' impression writings (details in Table~\ref{table:table2}). These terms appeared 47 times in the \textit{audio personas} group but only 9 times in the writings from the \textit{control} group. Aligning with the impression writings, participants were more likely to incorporate sounds into their impressions of others when sounds were anchored to the person (i.e., audio personas), compared to when sounds were anchored to objects.

Looking into the content of impression writings, the ~\textit{audio persona} group tended to describe how the sounds influenced their impressions of others. For instance, P2 described the research assistant with a positive audio persona: \textit{"I liked when he passed by me since the sound of the water gently splashing was nice and calming."} Similarly, P62 wrote ~\textit{"His sounds were like nature, which were complemented by his brown jumper and green hat. Together, that is peaceful"}. When recalling the person with a negative audio persona, P24 wrote:"\textit{He seems like a nice guy, although the buzzing sound wasn't as pleasant. The sound, along with his more defined facial structure like chin and cheekbones made him seem like a sharper guy who makes more judgments.}"  P26 wrote:"\textit{The clanking droid-like sounds inspires images of bluntness, stone faces, and assertiveness.}"

In the ~\textit{control} group, the impression writings were much more focused on the visual appearance of the research assistants. ~\textit{"He was also walking evenly and with purpose. Maybe he's chill, I might've gotten that impression from his clothes? He also might be interested in sports or fitness because he had a sports band and water bottle"}, wrote P12. 

\subsection{Finding \#2: Impact on Social Impression}

Figure~\ref{fig:study2_attract_threat} and Figure~\ref{fig:study2_sentiment} present the results related to social impressions. We used mixed ANOVA with ~\textit{audio anchoring} as a between-group factor and ~\textit{audio valence} as a within-group factor. This analysis was applied to measures including ratings of social attraction, threat, and likability as well as sentiment intensity in impression writing. While the two sounds within the valence condition have slight variation across the dependent variables, the trend remained the same and we opted to follow our preregistered plan of averaging across them within the condition, for ease of interpretation.

\subsubsection{\textbf{Social attraction.}} 
\begin{figure}[hbt!]
  \includegraphics[width=\linewidth]{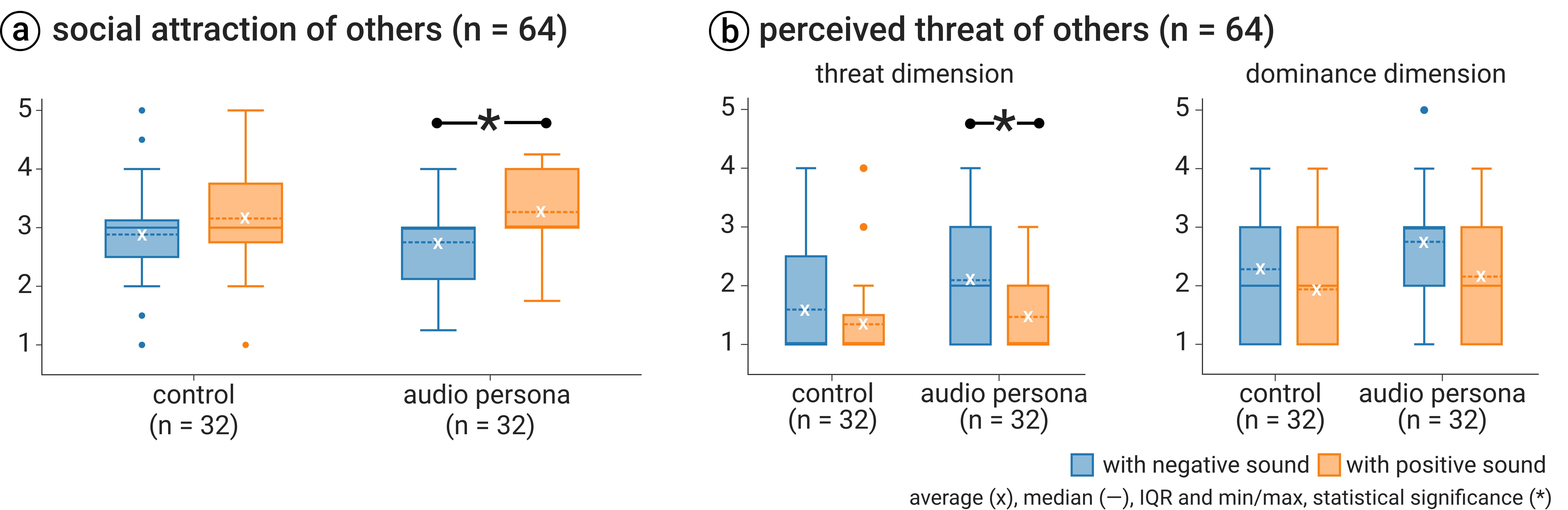}
  \caption{Self-report measures of (a) social attraction and (b) perceived threat of others. We reported the perceived threat construct by each item as the Spearman-Brown test indicated that the construct had a reliability lower than 0.7.}
  \Description{}
  \label{fig:study2_attract_threat}
\end{figure}

Figure ~\ref{fig:study2_attract_threat} presents the ratings for social attraction. A main effect was found for audio valence ($F(1,62) = 10.71, p = 0.002$), with a large effect size ($\eta^2 = 0.15$). No main effect was found for audio anchoring ($F(1,62) = 0.005, p = 0.94, \eta^2 = 0$) and no significant interaction was observed ($F(1,62) = 1.0, p = 0.32, \eta^2 = 0.016$).

While the interaction effect was not statistically significant as predicted (H4, H5), a significant main effect was observed for the audio valence. To identify the specific differences within each condition group regarding audio valence levels, as outlined in our preregistered hypotheses, we conducted exploratory pairwise comparisons, following prior work~\cite{kocur2021physiological, johnson2023unmapped,kirchner2024outplay}. Using Bonferroni-corrected pairwise T-tests, we found a significant difference ($p = 0.002$) in social attraction ratings within the \textit{audio persona} group. Participants rated the person with a positive audio persona ($M = 3.27, SD = 0.66$) as more socially attractive than one with a negative persona ($M = 2.75, SD = 0.75$), confirming \textbf{H2}.

In the \textit{control} group, the trend aligned with what we found in the ~\textit{audio persona} group, but it was not statistically significant ($p = 0.35$), failing to support H3. When sounds were anchored to objects, the person was rated slightly more socially attractive with a positive sound ($M = 3.16, SD = 0.81$) than with a negative sound ($M = 2.88, SD = 0.93$).

\subsubsection{\textbf{Perceived threat.}} 

On the threat dimension, a main effect was found for audio valence ($F(1,62) = 8.49, p = 0.005$) with a medium effect size ($\eta^2 = 0.12$). No main effect was found for audio anchoring ($F(1,62) = 3.53, p = 0.07, \eta^2 = 0.054$), and no interaction effect was observed ($F(1,62) = 1.56, p = 0.22, \eta^2 = 0.025$). As with social attraction ratings, we conducted exploratory pairwise comparisons to examine differences within main effects. Post-hoc pairwise comparisons showed a significant difference in threat perception ratings between valence levels in the \textit{audio persona} group ($p = 0.008$), with participants rating the person with a negative audio persona ($M = 2.09, SD = 1.13$) as more threatening than the person with a positive audio persona ($M = 1.47, SD = 0.71$). In the \textit{control} group, a similar trend appeared, though it was not significant ($p = 0.55$). When sounds were anchored to objects, the person was judged as slightly more threatening with a negative sound ($M = 1.59, SD = 0.93$) compared to the person with a positive sound ($M = 1.34, SD = 0.69$).

On the dominance dimension, a main effect was found for audio valence ($F(1,62) = 5.85, p = 0.02$), with a medium effect size ($\eta^2 = 0.086$). No main effect was found for audio anchoring ($F(1,62) = 3.35, p = 0.07, \eta^2 = 0.05$), nor the interaction effect ($F(1,62) = 0.42, p = 0.52, \eta^2 = 0.007$). Post-hoc pairwise comparisons showed no significant difference in dominance perception ratings between valence levels in the \textit{audio persona} group ($p = 0.056$) or \textit{control} group ($p = 0.49$). However, the trend still aligned with the threat perception ratings. In \textit{audio persona} group, participants rated the person with a negative audio persona ($M = 2.75, SD = 1.15$) as slightly more dominant than the person with a positive audio persona ($M = 2.16, SD = 1.06$). When sounds were anchored to objects, the person was judged as slightly more dominant with a negative sound ($M = 2.28, SD = 1.04$) compared to the person with a positive sound ($M = 1.94, SD = 1.0$).

\begin{figure}[hbt!]
  \includegraphics[width=\linewidth]{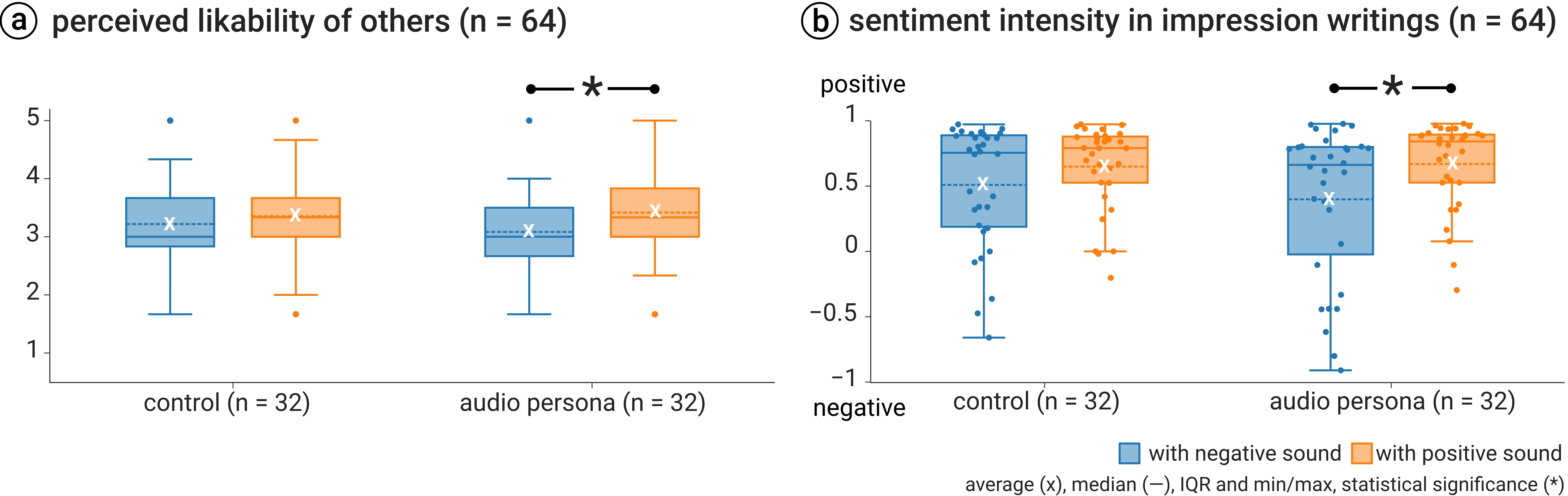}
  \caption{Self-report measures of (a) perceived likability of others and (b) sentiment intensity in impression writings. On a scale of -1 to 1, -1 represents the strongest negative sentiment, and 1 represents the strongest positive sentiment. }
  \Description{}
  \label{fig:study2_sentiment}
\end{figure}

\subsubsection{\textbf{Likability judgments.}} A main effect was found for audio valence ($F(1,62) = 6.50, p = 0.013$) on perceived likability, with a medium effect size ($\eta^2 = 0.095$). No main effect was found for audio anchoring ($F(1,62) = 0.06, p = 0.81, \eta^2 = 0.001$), and no interaction effect was observed ($F(1,62) = 1.16, p = 0.29, \eta^2 = 0.02$). Post-hoc pairwise comparisons showed a significant difference in likability perception ratings between valence levels in the \textit{audio persona} group ($p = 0.03$), with participants rating the person with a positive audio persona ($M = 3.42, SD = 0.71$) as more likeable than the person with a negative audio persona ($M = 3.08, SD = 0.69$). In the \textit{control} group, the trend aligned with ~\textit{audio persona} group, though it was not significant ($p = 0.6$). When sounds were anchored to objects, the person was judged as slightly more likable with a positive sound ($M = 3.35, SD = 0.68$) compared to a negative sound ($M = 3.22, SD = 0.73$).

\subsubsection{\textbf{Sentiments in impression writing.}}
As an exploratory measure, we conducted sentiment analysis on impression writings. We chose a lexicon-based model (NLTK Vader sentiment analyzer~\cite{hutto2014vader}), due to its emphasis on sentiment intensity, which provides a more granular and interpretable analysis compared to classification-based deep-learning models (such as~\cite{barbieri-etal-2020-tweeteval}). Despite its simplicity, we found it aligned well with human judgment. 

Sentiment intensity results are shown in Figure~\ref{fig:study2_sentiment}. We found a main effect for audio valence ($F(1, 62) = 9.12, p = 0.004$) on the sentiment intensity, with a medium effect size ($\eta^2 = 0.13$). No main effect was found for the audio anchoring ($F(1,62) = 0.27, p = 0.60$, $\eta^2 = 0.004$) and no interaction effect was observed ($F(1,62) = 0.91, p = 0.34$, $\eta^2 = 0.015$). 

We conducted exploratory post-hoc pairwise comparisons to reveal significance in different audio valence levels, as in the previous measures. We found that the sentiment was significantly more positive ($p = 0.035$) in impression writings describing the person with a positive audio persona ($M = 0.67, SD = 0.33$) compared to writings about the person with a negative audio persona ($M = 0.40, SD = 0.57$). ~\textit{Control} group had the same trend, though it was not significant ($p = 0.2$). When sounds were anchored to objects, the sentiment was slightly more positive in impression writings describing the person with a positive sound ($M = 0.65, SD = 0.32$) than those describing the person with a negative sound ($M = 0.51, SD = 0.46$).

\begin{figure}[hbt!]
  \includegraphics[width=0.5\linewidth]{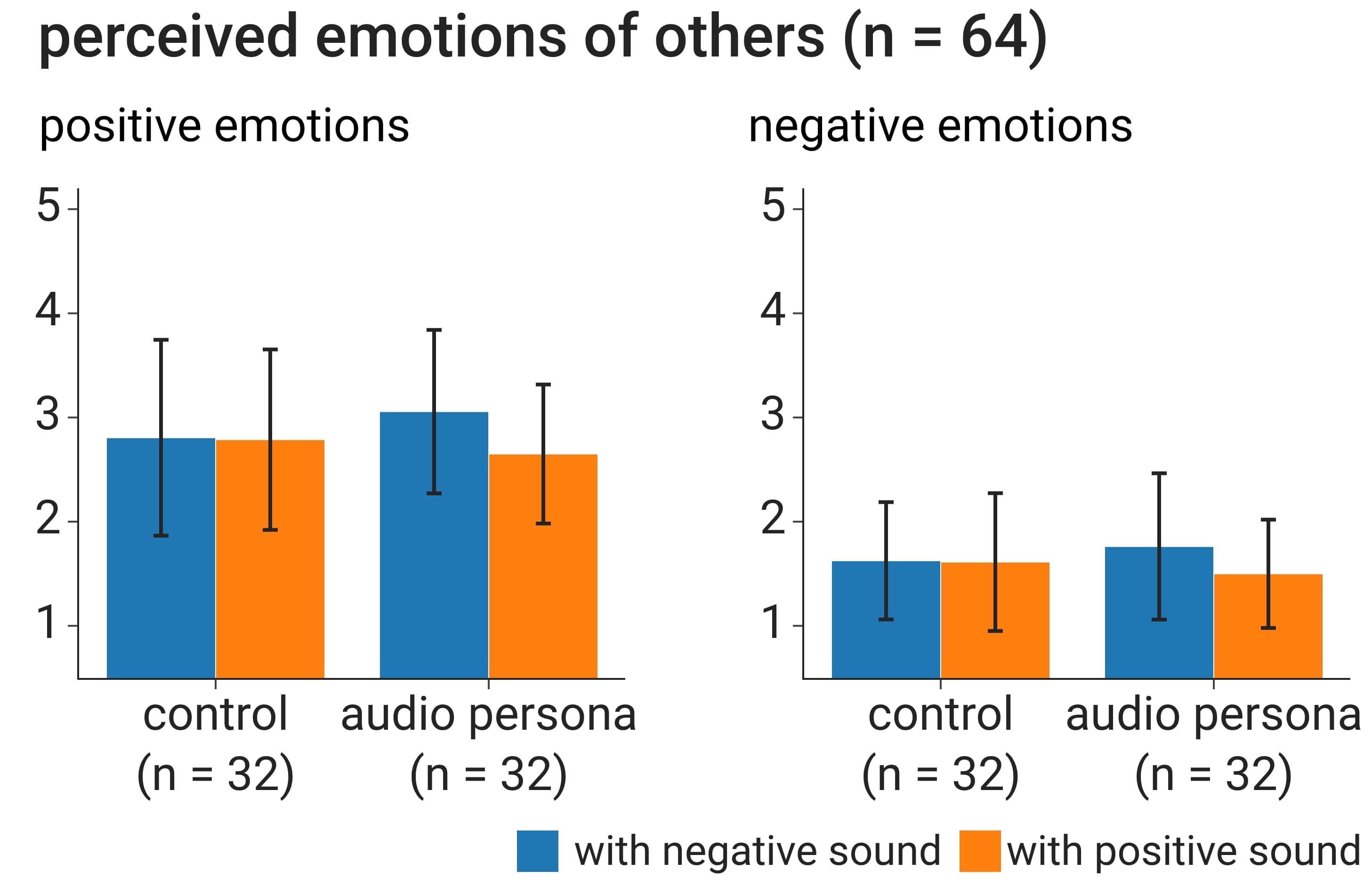}
  \caption{Self-report measure of perceived emotional states of others.}
  \Description{}
  \label{fig:study2_emo}
\end{figure}

\subsection{Finding \#3: Impact on Perceived Emotional States of Others}
Figure~\ref{fig:study2_emo} shows the results for perceived emotional states of others. The analysis revealed no significant main effects of either audio anchoring (F$(1, 62) = 0.10, p = 0.75, \eta^2 = 0.002$) or audio valence ($F(1, 62) = 3.60, p = 0.06, \eta^2 = 0.06$) on the perceived positive emotions in others. Similarly, no significant main effects was found for either audio anchoring ($F(1, 62) = 0.10, p = 0.92, \eta^2 = 0$) or audio valence ($F(1, 62) = 1.99, p = 0.16, \eta^2 = 0.03$) on the perceived negative emotions in others. No significant interaction was found between audio anchoring and audio valence for perceived positive emotions ($F(1,62) = 2.99, p = 0.09, \eta^2 = 0.05$) or perceived negative emotions ($F(1, 62) = 1.65, p = 0.20, \eta^2 = 0.03$). This indicates that participants generally evaluated the emotional states of both of their partners similarly, regardless of the audio valence or anchoring.

\subsection{Finding \#4: Impact on Social Behavior}
\textbf{Seat choice.} We first analyzed the seat picking choices at the end of the icebreaker task. In the \textit{audio persona} group, 14 out of 32 participants chose to sit with the research assistant that had a positive audio persona while 18 chose to sit with the research assistant with a negative audio persona. A chi-square test was conducted to examine the relationship between audio persona valence and choice of seat selection. The test statistic was ${\chi}^2(1, N = 32) = 0.5, p = 0.48$, indicating that there is no statistically significant relationship between the valence of the audio persona and the participants' choice of seating. Similarly, in the \textit{control} group, no significant relationship was found between the valence of sounds anchored to the objects objects and participants' final seat selections (${\chi}^2(1, N = 32) = 2.0, p = 0.16$). Specifically, 20 out of the 32 participants sat closer to the research assistant paired with a positive sound and 12 participants sat closer to the research assistant paired with a negative sound, with the sounds anchored to objects in the space. 

~\textbf{Minimum interpersonal distance.} Similarly, we analyzed the minimum interpersonal distance as participants approached each research assistant to show their answer sheets. This distance was calculated using the Euclidean distance between their head positions. Mixed ANOVA analysis revealed no significant main effects of either audio anchoring (F$(1, 62) = 0.21, p = 0.65, \eta^2 = 0.003$) or audio valence ($F(1, 62) = 0.09, p = 0.76, \eta^2 = 0.001$) on the minimum interpersonal distance. This suggests that participants generally kept a consistent distance from their partners, regardless of whether they were exposed to positive or negative sounds, and whether the sound was anchored to people or objects in the space.

\subsection{Summary of Findings in Study \#1}
Through a controlled in-lab study, we examined the effect of audio personas on social perception. Participants exposed to audio personas (i.e., body-anchored audio cues) actively integrated these sounds into their impressions of others, as evidenced by their impression writings and drawings. In contrast, participants in the control group—who heard the same sounds anchored to objects—were much less likely to associate the sounds with the person. This contrast highlights body-anchoring as a key design feature that enables audio personas to influence social perception. Furthermore, individuals with a positive audio persona were rated as more socially attractive, likable, and less threatening than those with a negative audio persona. Lastly, we found no significant effects of audio personas on perceived emotional states of others or on behavioral measures.
\section{Study \#2: User Scenarios and Designs of Audio Personas}

As Study \#1 demonstrated the effectiveness of audio personas in influencing social perception, Study \#2 explores potential use cases and design patterns for the concept. We conducted a design study with eight audio designers to uncover (1) social intentions behind using audio personas, (2) social contexts to use audio personas, and (3) design patterns of audio personas.

\subsection{Methods}
\subsubsection{\textbf{Participants}}
We chose to study participants with prior professional experience in audio design. This sample selection allows us to understand the potential contexts and intentions behind audio persona usage from a user perspective, as well as elicit designs with expert insights.

We recruited 8 participants with prior experience in producing/working with audio content from diverse backgrounds: audio engineering ($n = 3$), VR/AR design ($n = 2$), music ($n = 2$) and immersive arts ($n = 1$). Each participant had at least 2 years of experience in producing audio content and received training from university-level programs. The participants were recruited through word-of-mouth. 

\begin{figure}[hbt!]
  \includegraphics[width=\linewidth]{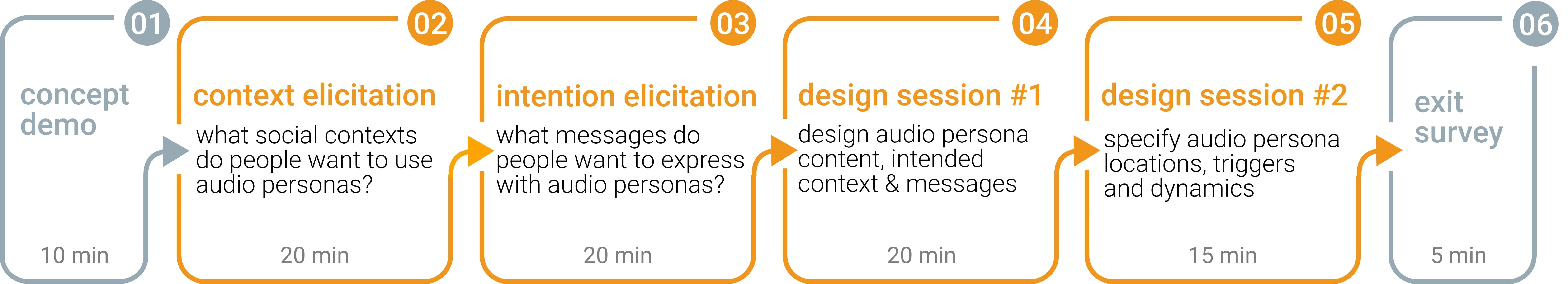}
  \vspace{-10px}
  \caption{Study \#2 procedure. In the first half of the study, participants tried out the device prototype and were interviewed about context and intentions for audio persona usage. In the second half of the study, participants created 6 different audio personas in two design sessions.}
  \Description{}
  \label{fig:study1_procedure}
  \vspace{-15px}
\end{figure}

\subsubsection{\textbf{Procedure}}
Participants were first introduced to the concept of audio personas. They put on the prototype system, as described in Sec. 3, and interacted with an experimenter to experience three examples of audio personas: (1) a bee sound when approaching the experimenter, (2) a "woosh" sound when the experimenter walked past, and (3) exaggerated stomping sounds from the experimenter. We chose these three demonstrations to show the technical possibilities of our prototype system, such as triggering audio personas based on proximity and anchoring sounds to body parts. These demos intended to provide inspiration, and participants were encouraged to think beyond these demo designs. 

We then interviewed participants about the social contexts where they believe audio personas could be useful. We began with open-ended questions and then discussed audio persona usage in different types of social contexts—\textit{public} (e.g., festivals, parks, streets, public transport), \textit{semi-public-private} (e.g.,  workspace, educational institutions, community events), and \textit{private} (e.g., home, private gathering)~\cite{karaccor2016public, swapan2019importance}. We also explored social messages suitable for audio personas to express. Again, we began with open-ended questions and discussed preferences for expressing specific types of messages using audio personas, including ~\textit{emotions}, ~\textit{ideas and opinions}, ~\textit{needs and desires}, ~\textit{traits and personality}, ~\textit{identity and belonging}, and ~\textit{personal experiences}~\cite{green2007self}.

Next, we elicited the creation of audio personas through two design sessions. Participants were asked to imagine themselves as designers of an audio persona library and create six different audio persona designs. In both sessions, participants used worksheets provided in the supplemental materials. In the first session, participants described what each audio persona should "sound" like, its social contexts, and intended message. In the second design session, they specified sound characteristics, including spatial locations, triggers, and dynamics. At the end of each session, participants reviewed and discussed their designs with the experimenters.

The study took approximately 75 - 90 minutes, and participants were compensated \$35 for their time. The study was approved by Stanford Institutional Review Board (IRB \# 71992).

\subsubsection{\textbf{Analysis}}
We video-recorded the sessions, transcribed the data, and organized it based on the guiding questions described previously. Two researchers were involved in the coding process. One researcher conducted an initial inductive thematic analysis of the transcripts to identify common themes~\cite{braun2006using}. The themes derived from the thematic analysis are reflected in the headings of the following subsections. Based on these themes, the two researchers independently coded each design created by the audio designers. They then discussed any discrepancies to reach consensus.

\subsection{\textbf{Finding \#1: Social Intentions of Using Audio Personas}}

 \begin{figure}[hbt!]
  \includegraphics[width=0.5\linewidth]{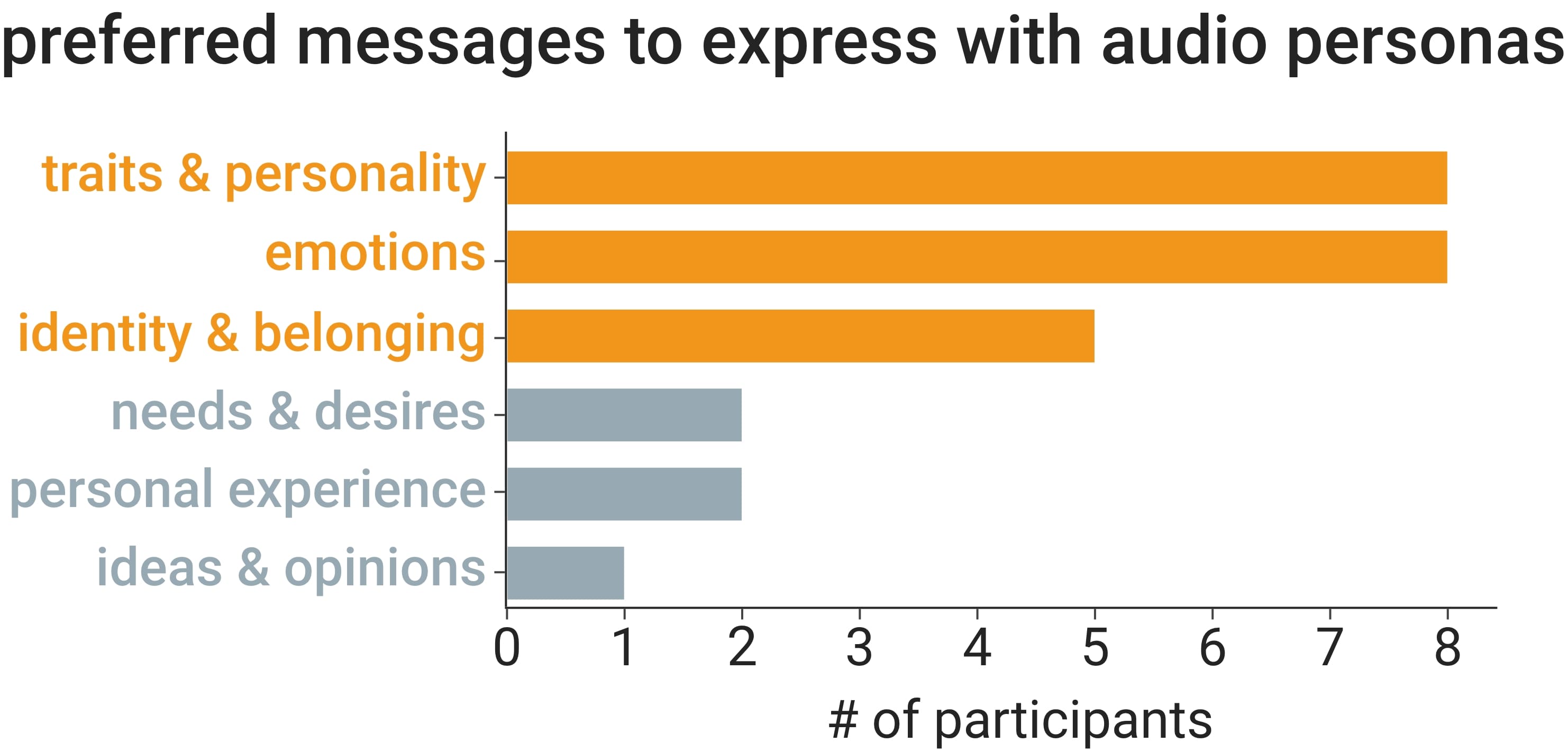}
  \caption{Participants' preferences for using audio personas to communicate different types of messages. Messages shown in orange represent those that over half of the participants found useful to communicate via audio personas. In contrast, messages shown in gray indicate those that participants felt audio personas were less effective or less helpful in communicating.}
  \vspace{-10px}
  \Description{}
  \label{fig:study1_msg}
\end{figure}

Figure~\ref{fig:study1_msg} shows participants' preferences for using audio personas to express  different types of social messages. All participants agreed that audio personas are well-suited for expressing "personality \& traits" and "emotions", followed by "identity \& belonging."  In contrast, only 2 out of 8 found audio personas suitable for expressing "needs \& desires" or "personal experience," with "ideas \& opinions" as the least favored.

Together with qualitative findings, we identified two major social intentions for audio persona usage: managing social impressions (e.g., "personality \& traits," "identity \& belonging") and signaling current states (e.g., "emotions"). We also identified the limitations of audio personas in conveying more complex information (e.g., "needs \& desires," "ideas \& opinions," "personal experiences").

\subsubsection{\textbf{Impression management}} When discussing their intentions of using audio personas, P2, P3, P4 and P5 drew parallels with the selection of clothing. Similar to clothing, audio personas offer "a way to express themselves" (P5), enabling others to "get that first impression based on something that the other person set" (P2). Even more, P3 elaborated that audio persona are able to convey messages difficult to express through clothing:

\begin{quote}
\textit{
"Somebody's audio persona would tell you another dimension of how they think~\dots some of them might be kind of cheeky or humorous and you can't necessarily or harder to convey with clothing. Unless you buy one of those T-shirts that have a joke on it; and those get really old really quickly."
}
\end{quote}

As a dynamic medium, audio personas could serve as \textit{"one of their intros when they're meeting people"} (P1), like a \textit{"business cards"} (P6),  showing \textit{"what they like, maybe what they don't like"} (P4), and fostering \textit{"affinities"} (P6). Moreover, audio personas were viewed as a means for creative expression. P8 described:

\begin{quote}
\textit{
"This is a nice way to kind of bring your personality to something more fictional, less rooted in reality~\dots you can really customize what other's perception of you is~\dots I feel like your physical presence feels different because of the sound."
}
\end{quote}

\begin{figure}[hbt!]
  \vspace{-10px}
  \includegraphics[width=\linewidth]{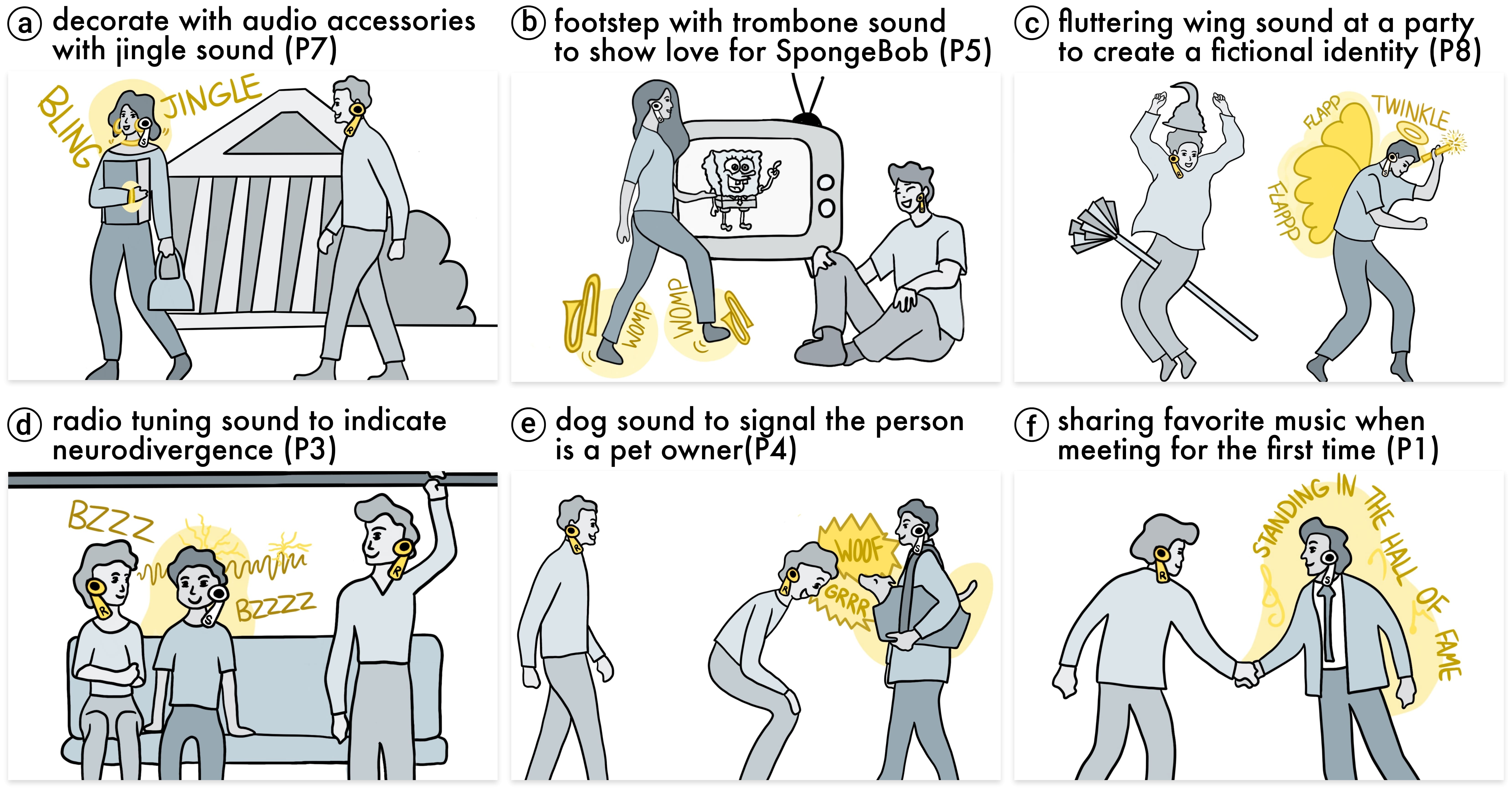}
  \vspace{-20px}
  \caption{Illustrations of sample audio persona designs from Study \#2, with the intention of managing impressions. The authors of this paper visualized the audio persona design based on information provided by the participants. The yellow highlights indicate the content of the audio persona.}
  \vspace{-10px}
  \Description{}
  \label{fig:study1_design1}
\end{figure}

In the designs created by participants, 21 out of 48 audio personas intended to manage impressions, focusing on influencing how others perceive users. A wide range of audio content with varying valence was used and Figure~\ref{fig:study1_design1} shows a selection of them. For instance, P7 designed an audio persona that features auditory accessories with a jingling bell sound to reflect personal jewelry taste and personality, as shown in (Figure~\ref{fig:study1_design1}a). P5 chose to augment their footstep sounds with a trombone sound, showing the love of SpongeBob (Figure~\ref{fig:study1_design1}b). P4 designed an audio persona that emits a dog barking sound whenever they carry their dog in a pouch, signaling at a distance that they are a pet owner and cuing others to notice the dog's presence (Figure~\ref{fig:study1_design1}e).

\subsubsection{\textbf{Signaling current states}} 
Another social intention for using audio personas is to signal the users' current states and inform others on how they prefer to be interacted with. Audio personas offer an "implicit" channel~\cite{weiser1996designing} to convey current states, without saying out loud and \textit{"imposing too much on [others]"} (P4). P1 described it as \textit{"another social cue in terms of emotions"} and P5 saw it similar to \textit{"setting your status"} commonly found in chat apps. 

P7 described examples of how they might use audio personas to signal current states. Different valences of sounds were leveraged to signal different emotional states.

\begin{quote}
\textit{"If I'm not feeling super chatty and I don't want to engage in conversation with them, the [audio persona] can be something softer or more melodic, less exciting, more just like a kind of monotone. Versus like, if I'm feeling really high energy, it can be a more high energy sound"}, said P7. 
\end{quote}

Such indications of one's current state could help others to \textit{"empathize better with where one's at"} (P3) and \textit{"help other people to know how to interact with you"} (P8). For instance, P2 described:

\begin{quote}
\textit{
"If someone's like, kind of down or depressed, like, we hear the rainfall?... maybe we could like, approach them and be like, `Oh, are you okay?'"
}
\end{quote}

\begin{figure}[hbt!]
  \includegraphics[width=\linewidth]{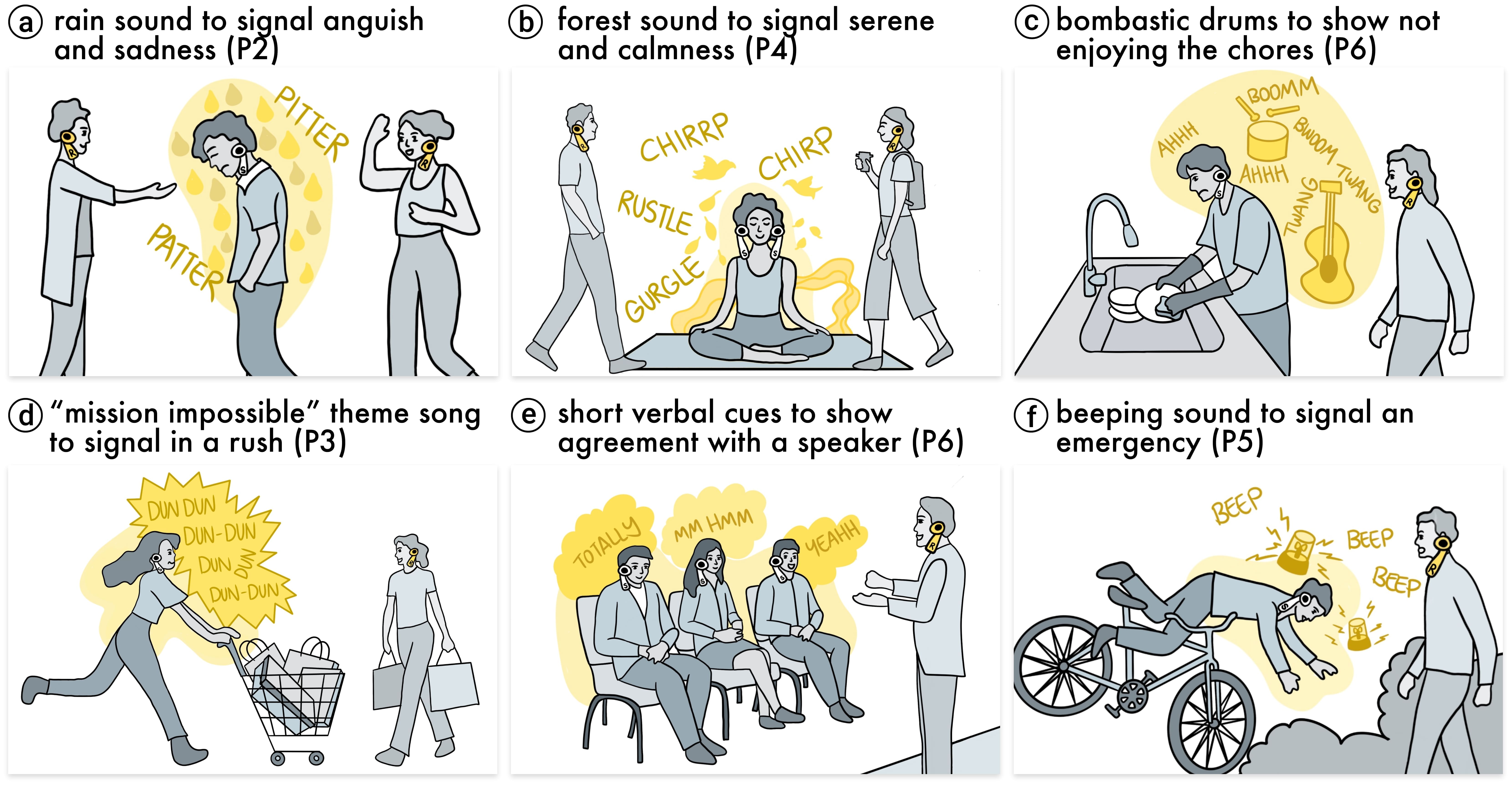}
  \caption{Illustrations of sample audio persona designs from Study \#2, with the intention of signaling current states. The authors of this paper visualized the audio persona design based on information provided by the participants. The yellow highlights indicate the content of the audio persona.}
  \Description{}
  \label{fig:study1_design2}
\end{figure}

21 out of 48 designs were created to signal current states, reflecting users' emotions, needs, or intended interactions, and Figure~\ref{fig:study1_design2} shows a selection of them. For instance, P3 used the \textit{Mission Impossible} theme song to signal urgency in settings like a grocery store, prompting others to step aside or assist (Figure~\ref{fig:study1_design2}d). In contrast, P4 designed peaceful forest sounds to indicate serenity and a desire not to be disturbed during activities like meditation (Figure~\ref{fig:study1_design2}b). P6 created a bombastic drum and metal guitar persona to express their dislike for chores such as dishwashing (Figure~\ref{fig:study1_design2}c).

\subsubsection{\textbf{Limitations in delivering complex information}} When discussing messages that audio personas might not be suitable to convey, participants referred to those involving complex information or narratives, such as opinions and personal experiences. P2 attributed the limitations to how audio is prone to \textit{"perceptual bias"} and P5 stated these information need \textit{"more clarity"}. For instance, when considering ideas and opinions, P8 described them as \textit{"very nuanced"}. \textit{"While audio is nuanced, it is always subjective. And so I don't think it would clearly convey whatever idea or opinion would want it to pitch or rather, it has the potential to be misinterpreted"}, said P8. Similarly, P4 expressed that personal experience is \textit{"too complex a subject almost to capture in an audio clip"}:

\begin{quote}
\textit{
"That's like you have to imply something that is reminiscent of your childhood or reminiscent of a recent life experience and hope that people capture that implied message; but I don't think it'll be as effective as compared to some of the other options."
}
\end{quote}

 P1 suggested a good audio persona should be~\textit{"something that is not too long, like something that's iconic, and then you could identify quickly."}

\subsubsection{\textbf{Additional use cases for audio personas.}} 
Six designs did not fall into impression management or signaling current states. Two were situated in motor learning contexts. For instance, P05 designed smoothing and relaxing sounds that played when the teacher's and student's body motions were aligned. In this case, the audio served as immediate feedback audible to both the sender and receiver. Two designs leveraged audio personas to evoke or recall past events. For example, P08 used the sound of ocean waves while retelling a vacation experience, immersing the listeners in the story. We also observed audio persona use in accessibility contexts. By listening to changes in one's own audio personas during locomotion (P08), blind and low vision individuals could gain spatial awareness of the space. Last but not least, audio personas were used as motivational social gestures, allowing users to send encouragement—such as the message “you got this”—to others.

\subsection{\textbf{Finding \#2: Social Context for Using Audio Personas}}

\begin{figure}[hbt!]
\includegraphics[width=0.5\linewidth]{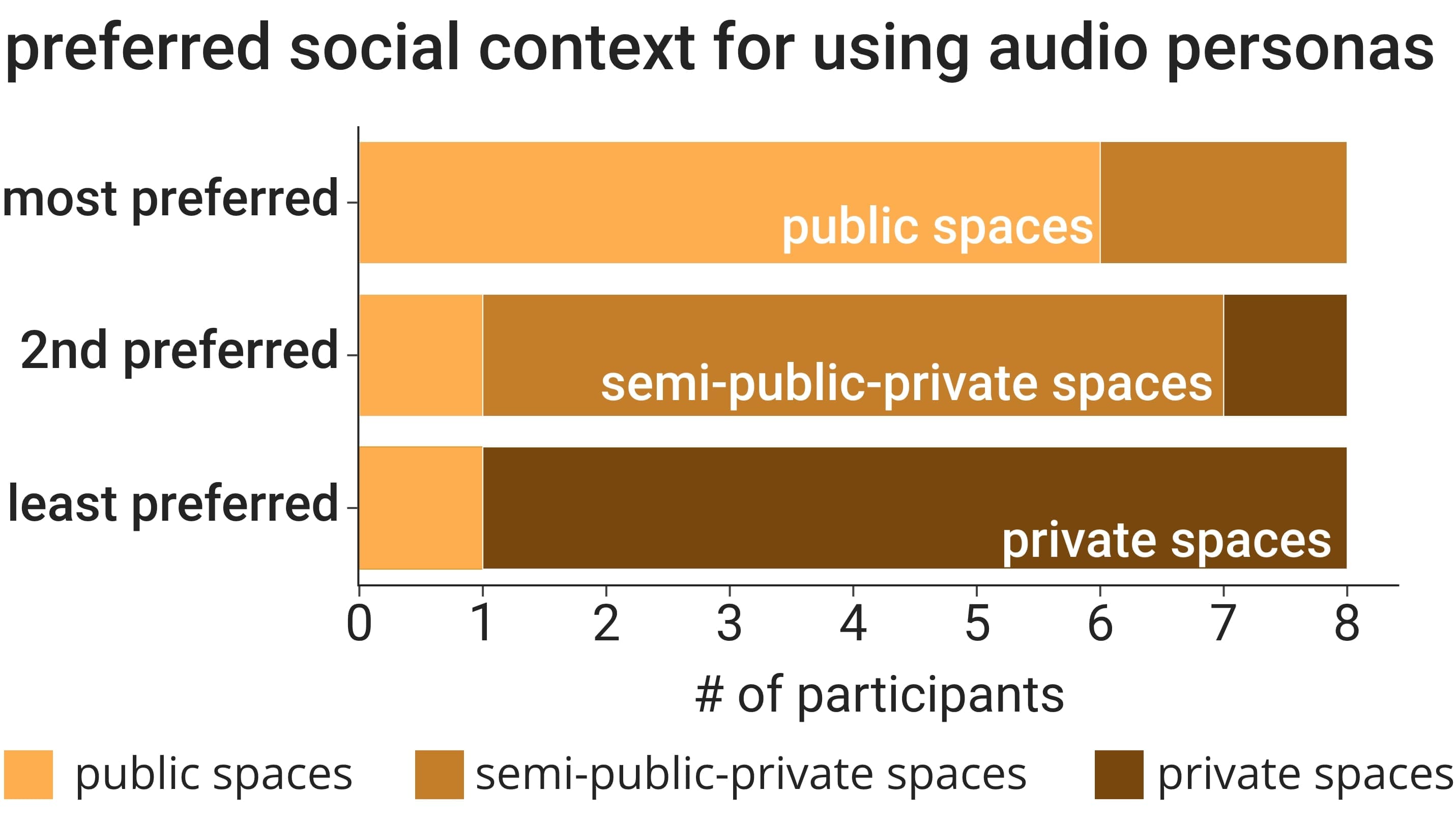}
  \vspace{-5pt}
  \caption{Ranking of preferred social context to use audio personas. Public spaces were found to be the most preferred, followed by semi-public-private spaces. Participants found private spaces to be least suitable for using audio personas.}
  \Description{}
  \label{fig:study1_context}
\end{figure}

Figure~\ref{fig:study1_context} shows the participants' rankings of their preferred social context for using audio personas. Public spaces (e.g., festivals, parks, streets, public transport) and semi-public-private spaces (e.g., workspace, educational institutions, community events) were found to be most preferred. Specifically, 6 out of 8 participants rated public spaces as where audio personas will be most useful for, followed by semi-public-private spaces. 7 out of 8 participants found the private space to be the least useful for using audio personas. In terms of audio persona designs, 36 designs were created for semi-public-private environments, 14 for public spaces, and 5 for private settings. The total exceeds 48 because some designs are intended for multiple contexts.

\subsubsection{\textbf{Intended for public and semi-public-private spaces}}
In a public context, P6 described their experience of communications as:\textit{"you're not actually able to verbalize your thoughts to as many people as you want as comfortably"}. Thus, audio personas can provide a nice way to \textit{"get first impressions of strangers"} (P2), serve as \textit{"a novel way to meet and greet"} (P3) and \textit{"communicate some sort of affinity"} (P6). Thus, users can potentially \textit{"make new connections"} as audio personas become something that people can \textit{"bond over"} (P4). P3 echoed, \textit{"In public contexts, audio personas could help you network with new people that are similar to you"}. 

In semi-public-private spaces, P4 described the social interactions as: \textit{"there are still gonna be a lot of strangers, and potentially coworkers."} Similarly, P2 described those spaces as more "familiar" and they would \textit{"become more open to different [audio] personas"}. Similar to its use in public spaces, audio personas can serve as \textit{"an interesting icebreaker~\dots to get to know somebody else."} (P8). The usage of audio personas in semi-public-private space also involve \textit{"understanding the vibes of somebody a little bit better"} (P3). They elaborated with an example: \textit{"When somebody is really zoned into something, they don't want to be bothered at all~\dots I can imagine, audio personas~\dots being used for things like personal space violation."}
 
\subsubsection{\textbf{Less useful in private space}}
When considering a private social context, participants found it to be the least suitable for using audio personas. P1 described private gatherings as being:~\textit{"already pretty small. [and people are] in the same space and talking with each other." } P3 added that in those contexts, people usually ~\textit{"know each other well enough"}. Thus, participants deemed people should be able to "start up a conversation" (P2), ~\textit{"just talk to them honestly"} (P4) and ~\textit{"be fully present engaging with the person"} (P8). In these contexts, audio personas were described as "kind of unnecessary complication and distraction" (P3), making them feel ~\textit{"less personal"} (P5) and ~\textit{"more disconnected from the people around them for conversation"} (P7). On the other hand, P8 and P6 also described scenarios where they might want to use audio personas in a private context. P8 said:~\textit{"If I want to use it in a private space, it's more of a novelty."} P6 envisioned an asynchronous usage where users re-experience other's audio personas after private gatherings.

\subsection{Finding \#3: Design Patterns of Audio Personas}

Lastly, we analyzed design patterns in the sound characteristics of audio persona designs, specifically examining their localization, activation triggers, and dynamic properties.

\subsubsection{\textbf{Localizations}}

\begin{figure}[hbt!]
  \includegraphics[width=\linewidth]{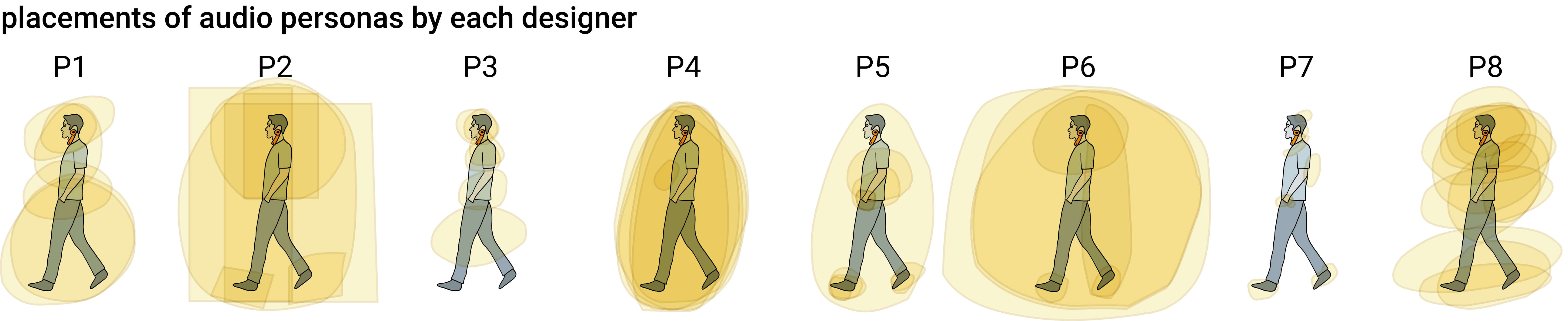}
  \vspace{-10pt}
  \caption{Visualizations of the spatial locations of audio personas designed by each participant. Each subgraph shows the overlaid spatial locations that participants drew for their designs. Darker areas on the figure indicate locations where more designs are intended to anchor an audio persona to.}
  \Description{}
  \label{fig:study1_locations}
\end{figure}

We traced the sketches of audio persona locations marked by participants on the design worksheets and overlaid them to identify design patterns, as shown in Figure~\ref{fig:study1_locations}. There was variability in the specific body parts where the audio personas were attached. 25 out of 48 designs were anchored to the upper body of the users, such as to the head (n = 11), hands (n = 4), and back (n = 3). 15 designs were placed around the lower body of the users, predominantly on the feet (n = 11). In 12 of 48 designs, participants circled the whole body instead of specifying a center origin, indicating they intended the audio personas to be heard around the body without a precise origin. 8 designs placed audio personas on multiple body parts.

While most designs were intended for individual use, P6 and P8 suggested that multiple people could share a group audio persona when engaging in similar activities, with each person contributing a part. In these cases, the audio persona would be placed at the group's location, rather than from a single individual.

\subsubsection{\textbf{Triggers}}
\begin{figure}[hbt!]
  \includegraphics[width=\linewidth]{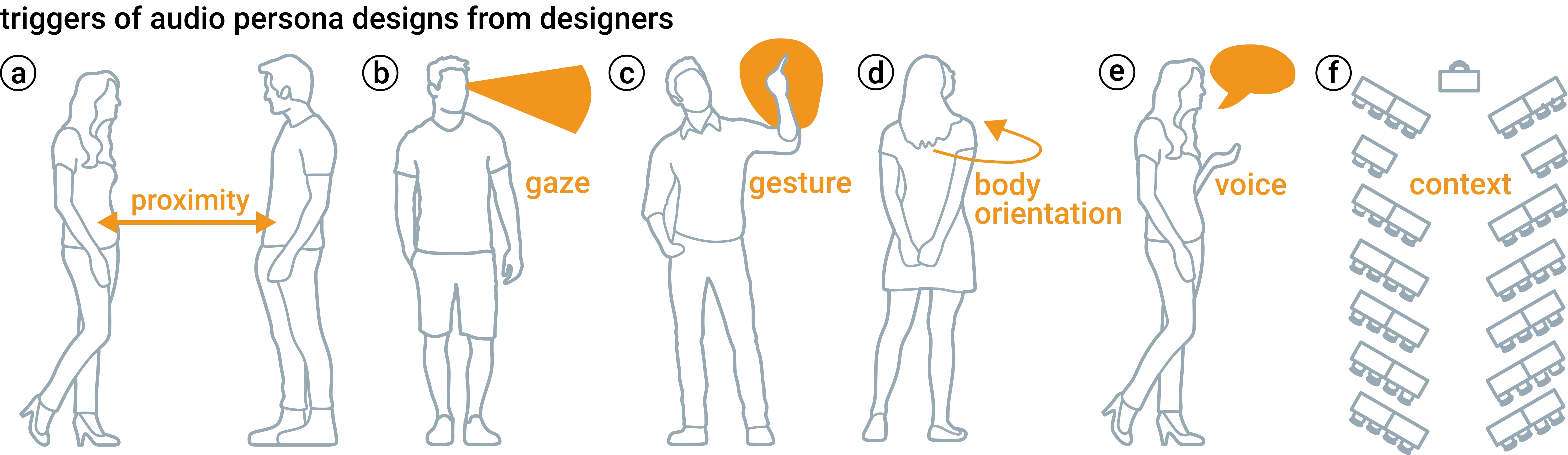}
  \caption{Triggers for audio personas used in designs created by participants. These triggers could be used individually or combined. (a) proximity, (b) gaze, (c) gestures, (d) body orientation, (e) voice and (f) context.}
  \Description{}
  \label{fig:study1_trigger}
\end{figure}

Regarding what triggers the start of audio personas (i.e., when audio personas become audible to others), proximity is the most commonly used, as shown in  Figure~\ref{fig:study1_trigger}. Out of 48 designs, 30 were set to trigger when the distance between two users reached a predefined threshold. While most designs were triggered when others approached, in one of P6's designs, they chose to trigger the audio persona only when someone walked past the user, calling it "exit music." 7 designs used specific gestures to trigger actions, such as the "OK" hand gesture (P1) or air drumming motions (P7). 2 designs relied on gaze, like eye contact, as the trigger. 8 designs were context-dependent, activating in specific situations or when users engaged in certain activities, such as in a lecture hall (P6) or when a dog was nearby (P4). Additionally, participants explored using voice, body orientation (e.g., the direction the user is facing), or manual activation by the user. 8 designs explored a combination of multiple triggers.

\subsubsection{\textbf{Dynamics}} 
Most audio personas designs from Study \# 1 are dynamic, with evolving sound characteristics. 11 out of 48 designs involve a complete change in audio content based on factors like proximity, gestures, or user preferences. In one of P1's designs, the audio persona shares the user's favorite music, which adapts to the user's changing taste. Other changes in dynamics involve the volume (n = 9), length (n = 7), pitch (n = 3), and speed (n = 2) of audio. For instance, in P2's design, rain sounds vary in intensity to reflect the mood, becoming lighter as the person feels better. In 10 out of the 48 designs, participants chose to keep the audio persona unchanged.

When multiple people are present, the blending of their audio personas become possible. For example, in P8's design, a group shares an audio persona (e.g., forest or river sounds) during a group meditation to create a collective sense of calm. Each user retains their own sound and, as more people join, additional sounds, like bird sounds, are introduced.

\subsection{Summary of Findings in Study \#2}
In Study \#2, we explored social context and intended social messages for using audio personas by working with participants {experienced in audio design. We identified potential use cases in managing social impressions (e.g., personality \& traits, identity \& belonging) and signaling current states (e.g., emotions) but are limited in communicating complex information (e.g., ideas \& opinions). Participants preferred using audio personas in public (e.g., street, transport) and semi-private settings (e.g., workspace, educational institutions) to form bonds with people they do not know well, but found it less useful in private spaces. Analysis of 48 participant-generated audio personas revealed that a wide range of audio content with varied valence was used, and there are diverse design possibilities for localization, triggers, and dynamics.
\section{Discussion}
In this section, we discuss insights from our studies, limitations of this paper, and future directions for audio personas.

\subsection{Sound, Impression Management and Social Perception}
Study \#1 employed an in-lab, controlled study to examine the effectiveness of audio personas, body-anchored sounds, in influencing social perception. 

Our findings showed that participants integrate audio personas into their impressions of others, revealed in the impression writings and drawings. In the \textit{control} group, where participants heard the same sounds anchored to objects, it was much less likely to integrate sounds into their impressions of others. These findings align with a theoretical framework of impression formation~\cite{livesley1973person, workman1991role}, which suggests that individuals form impressions by focusing on specific cues from the available information about a person and making inferences based on those cues. From a multisensory cue combination standpoint~\cite{spence2013just, zaki2013cue}, an audio persona provides both spatial and temporal alignment between an auditory cue and a person, strengthening the association of sound as available information about the person. In contrast, in the \textit{control} group, sounds anchored to objects had only temporal alignment (i.e., the sound played when the person was present) but lacked spatial alignment with the person. Such a multisensory integration process could explain impression formation in the context of audio personas.

Study \#1 also investigated the emotional component of audio persona. We found participants rated the person with a positive audio persona as more socially attractive, likable, and less threatening than the person with a negative audio persona. We observed a similar trend in the ~\textit{control} group, where sound was anchored to objects during interactions with others, although the difference was not statistically significant. These results aligned with prior work that showed emotion-evoking music can influence social impressions of individuals in photographs in the same direction of the music valence~\cite{may1980effects, hanser2015effects}. 

We did not observe a significant interaction between audio anchoring and audio valence as originally hypothesized. We explicitly chose not to use a control group with no sound in order to measure the clear effect of audio anchoring. The subtlety of the manipulation suggests that the mere presence and temporal binding of audio may elicit a similar level of affective response, regardless of where the sound is spatially anchored. However, body-anchored sound does enhance the conscious integration of sound into impressions of the person, as demonstrated in the impression writings and drawings. This distinction may become especially important in multi-person interaction contexts, where individuals engage with several people simultaneously, unlike the one-on-one interactions explored in this paper. In such settings, body-anchored sounds may make a clearer distinction, potentially leading to more differentiated social impressions of each person, compared to situations where multiple sounds are simultaneously present in the space.

While previous research suggested that sounds prime the perceived emotions of others~\cite{logeswaran2009crossmodal, ignacio2021music, carroll2005priming}, our study did not observe a similar effect. One possible explanation is that our study put participants into more complex social settings (beyond observing static faces as in prior work), where interpreting perceived emotions of others involves processing other cues~\cite{lange2022reading}. Since most non-verbal cues were controlled and kept neutral in our study, participants likely perceived the primary signals as neutral, overriding the priming effect of audio. This indicates that semantic congruence~\cite{laurienti2004semantic, kim2022studying} between audio personas and other contextual information might be necessary for influencing the perceived emotional states of others.

We did not observe main effects on certain  behavioral measures, such as seat choice and interpersonal distance. These results are partially consistent with prior research on how sound influences interpersonal distance. While previous studies have found effects on audio in interpersonal distance, these were typically observed in passive approach tasks—where the experimenter approached the participant, and the participant indicated when to stop~\cite{tajadura2011space}. In contrast, no effects were found in conditions where participants actively approached the experimenter~\cite{tajadura2011space}. In our study, participants actively chose where to position themselves, which may explain the lack of significant behavioral effects and need further investigation.

\subsection{Design Considerations for Audio Personas}
Study \#2 involved audio designers to elicit potential use cases and design patterns for audio personas. Based on these findings, we synthesized following design considerations. 

\textbf{Designing for shared spaces.} Considering social contexts to use audio personas, participants in our design study preferred using audio personas in public (e.g., street, transport) and semi-private settings (e.g., workspace, educational institutions) to form bonds with people they do not know well. As audio personas are envisioned for these shared spaces, a key design challenge is how to effectively manage the projection and reception of multiple audio personas. It’s easy to imagine situations where hearing multiple audio personas simultaneously could become distracting or overwhelming. One potential solution is to draw inspiration from online social media platforms, where users manage their networks by choosing who to subscribe or unsubscribe. Similarly, audio persona systems could give users control over who can hear their audio persona and whose personas they want to receive. Additionally, as suggested by several designs in from Study \#2, blending or mixing multiple personas into a single ambient layer could help users sense the overall "vibe" of the space without being overwhelmed by different sounds. Users could then selectively tune into a specific person’s audio persona if desired.

\textbf{Privacy, security and ethical considerations.} Considering the social intentions behind using audio personas, participants in our design study expressed a preference for using them to manage social impressions (e.g., personality \& traits, identity \& belonging) and signaling current states (e.g., emotions). Building upon these use cases, security, privacy and ethical considerations need to be taken into account. 

Given that audio personas may contain personal information intended for others, privacy protections are essential. Users should have full control over with whom their audio personas are shared. It is also important that users are made aware of when their audio personas are being heard during interactions—for example, through notifications. Additionally, because audio personas are tied to a user’s spatial location, continuous broadcasting may unintentionally reveal behavioral patterns that users did not intend to share. To mitigate this risk, designers might consider suggesting spatial and temporal limits—such as setting the maximum distance within which an audio persona can be heard, or maximum duration it remains active.

Designers should also be mindful of the potential harms associated with misuse of audio personas and develop appropriate intervention strategies. For instance, malicious actors could exploit the system by broadcasting loud, overwhelming sounds to capture attention or by transmitting inappropriate audio content. To mitigate such risks, one approach is to implement content moderation mechanisms, such as pre-filtering, to block disruptive or harmful audio before it reaches listeners. Additionally, in cases where audio personas are also audible to the senders themselves, potential effects on gait and movement should be considered, as highlighted by prior work on movement sonification~\cite{d2024soniweight, tajadura2015action}.

Last but not least, we recognize that the value of audio personas depend on how widely they are adopted. We also acknowledge that the current headphone-based prototype poses a technology barrier, limiting the augmented social experience to users with headphones. However, audio personas do not have to be a headphone-delivered experience. For example, personal speaker-based (e.g., "boomboxes,"~\cite{boyer2014urban} popular in the 1980s and 90s) or clothing-integrated audio personas could achieve more universally accessible experiences. 

\textbf{Individual difference to audio stimuli.} Analysis of 48 participant-generated designs indicates that audio personas offer rich design possibilities. Participants explored a wide range of audio content with varied emotional valence and highlighted diverse patterns for localizations, triggers, and dynamics. As designers continue to explore the design space of audio personas, one consideration to be noted is individual differences to audio stimuli. For instance, certain sounds might be deemed positive by some people while deemed as negative by others. Thus, communication through audio personas may break down if senders and receivers hold different mental models of the same sounds. A potential solution is to learn from usage patterns and construct an audio persona library that includes designs with commonly understood meanings. This is similar to emoji libraries commonly seen on social media platforms. Building on such a shared audio persona library, designers could then offer customization options to support personalized expression.

\subsection{Limitations}
This paper's investigations serve as key initial steps toward exploring audio personas but also come with several limitations.

On the system side, the current prototype relies on base stations and optical trackers. While it was sufficient for in-lab investigations of the concept, the existing system restricts interactions to equipped spaces and users wearing them. A more portable, user-friendly form factor is needed for real-world scalability. Future work could explore ultra-wideband (UWB) or computer vision for spatial tracking and leverage mobile devices for audio output. These improvements would enhance the accessibility and practicality of audio persona systems in everyday settings.

Study \#1 employed a controlled, in-lab experiment to investigate whether audio personas social perception. While this design provided valuable insights into how key design factors of audio personas (e.g., body-anchored, audio valence) influence social perception, it has a few limitations. First, the behavior of research assistants was pre-scripted, which ensured experimental control, but did not fully capture the nuance of everyday interactions. Future work should examine audio persona use in more socially complex settings. For instance, researchers can incorporate other communication channels (e.g., spontaneous conversations) to explore how audio personas influence impression formation with additional contextual cues. Similarly, researchers should investigate the effects of audio personas without the explicit framing of "studying how audio impacts social interactions" to encourage naturalistic reactions. 

Second, the study used object-anchored sounds as the control condition, which allows us to examine the exact effect of body-anchoring in audio personas. However, future work should compare audio persona with other baseline, such as no audio or non-spatialized sound. Third, in Study \#1, we used only nature and animal sounds, but a wider variety of content should be explored, such as human and mechanical sounds. This study also only focused on a single style of sound characteristics, investigating variations in localization, triggers, and dynamics of audio personas will provide new insights of how audio personas influence social perception. Studying a broader range of audio persona designs will also help us to understand their differences in affecting social perception. Ultimately, the effects of audio personas in multi-user contexts and real-world social interactions beyond the lab should be explored.

Study \#2 interviewed participants experienced in audio design to gain initial insights of audio personas usage, following prior work~\cite{liu2023visual}. A larger and more diverse sample, particularly participants without a technical background, is needed to understand the preferences of the broader population. Additionally, while Study \#2 elicited an initial set of designs, it did not fully explore the challenges of scaling audio personas, such as in large conference settings. Future studies should invite designers to investigate techniques like audio mixing for effective deployment in large-scale, multi-user contexts.

\subsection{Future Directions}
To further explore the potential of audio personas, future research should investigate their use in more naturalistic settings. This would extend the findings from our in-lab study and build toward the potential use cases identified in the design study. One promising direction is the deployment of audio persona systems at conferences. Conferences are semi-public-private environments where attendees often seek to connect with new people over the course of a few days. Such settings offer a unique opportunity to explore the use cases and social functions of audio personas in a scalable yet manageable way. Moreover, they provide a valuable platform for studying how individuals manage multiple audio personas to present themselves differently across varying contexts \cite{tajfel1974social}, as well as for exploring design strategies to support hearing multiple audio personas in shared spaces.
 
As audio personas move beyond in-lab environments into more naturalistic settings, another direction for future work lies in exploring the psychological dynamics underlying their use. In the in-lab study, we were able to control the behavior and facial expressions of research assistants to remain neutral, isolating the effect of audio personas. However, in real-world scenarios, a person’s facial expressions and body movements are dynamic and may not always align with the tone conveyed by their audio persona. For example, someone might display a cheerful facial expression while their audio persona sounds sad. This kind of mismatch raises important research questions about the underlying processes of social perception when additional—and potentially conflicting—cues are introduced.

As future work explores more complex audio personas, leveraging knowledge in psychoacoustics offers great potential. For example, the best human auditory localization performance occurs within \ang{1}-\ang{2} in the horizontal plane and \ang{4}-\ang{5} in elevation~\cite{carlini2024auditory}. In Study \#1, we observed that audio designers localized audio personas to specific body parts to create motion-specific experiences, aligning well with these perceptual thresholds. Further research could investigate computational models to support the creation of perceptually distinct audio personas. This becomes particularly important to prevent sensory confusion in multi-user contexts.

\section{Conclusion}
We introduced Audio Personas, a novel concept to augment face-to-face interactions with body-anchored audio cues. Our in-lab study revealed that audio personas can influence how people formed social impressions. Individuals with a positive audio persona were deemed as more socially attractive, likable, and less threatening than ones with a negative audio persona. Through a design study with audio designers, we further uncovered potential use cases and elicited design patterns. Audio personas open new possibilities for engaging dynamic multisensory channels to enhance face-to-face social interactions and this paper laid out key building blocks for future explorations.

\begin{acks}
We would like to thank Keshav Rastogi, Muhammad Ali, and Brian Beams for their support on the user study logistics. We thank Anna Matsumoto and Emmanuel Corona-Moreno for their help on the paper video. We would also like to thank everyone who provided feedback on this project: Ethan Buck, Prof. Ana Tajadura-Jiménez, Danyang Fan, Han Guo, Eugy Han, Saehui Hwang, Monique Santoso, Tara Srirangarajan, Olivia Tomassetti, Ahad Rauf, Savannah Cofer, Portia Wang, Yifei Cheng, Yiran Zhao, and Ran Zhou. The lead author of this work is supported by the Knight-Hennessy Scholarship.

\end{acks}

\bibliographystyle{ACM-Reference-Format}
\bibliography{reference.bib}

\end{document}